\documentclass[12pt,subeqn,a4paper]{article}

\usepackage{amssymb,amsmath,amsfonts,amsthm, amscd, mathrsfs,helvet}
\usepackage{bm}
\usepackage{graphicx,verbatim,color}
\usepackage{longtable}
\usepackage{psfrag}
\usepackage[all]{xy}

\theoremstyle{plain}

\numberwithin{equation}{section}

\begin{document}

\begin{titlepage}
\begin{flushright}
{\bf March 2007} \\
DAMTP-\\

hep-th/yymmnnn \\
\end{flushright}
\begin{centering}
\vspace{.2in}
{\large {\bf Giant Magnons and Singular Curves}}\\

\vspace{.3in}

Beno\^{\i}t Vicedo\\
\vspace{.1 in}
DAMTP, Centre for Mathematical Sciences \\
University of Cambridge, Wilberforce Road \\
Cambridge CB3 0WA, UK \\
\vspace{.2in}
%
%
\vspace{.4in}
{\bf Abstract} \\

\end{centering}

We obtain the giant magnon of Hofman-Maldacena and its dyonic
generalisation on $\mathbb{R} \times S^3 \subset AdS_5 \times S^5$
from the general elliptic finite-gap solution by degenerating its
elliptic spectral curve into a singular curve. This alternate
description of giant magnons as finite-gap solutions associated to
singular curves is related through a symplectic transformation to
their already established description in terms of condensate cuts
developed in hep-th/0606145.

\end{titlepage}



\input{epsf}

\setcounter{section}{-1}

\section{Introduction}

Recently, a certain limit of the AdS/CFT correspondence was
proposed by Hofman and Maldacena \cite{Hofman:2006xt} in which the
't Hooft coupling $\lambda$ is held fixed allowing for a direct
interpolation between the gauge theory ($\lambda \ll 1$) and
string theory ($\lambda \gg 1$). In the Hofman-Maldacena (HM) limit,
the energy $E$ (or conformal dimension $\Delta = E$) and a $U(1)$
R-charge $J_1$ both become infinite with the difference $E - J_1$
held fixed. On the string side, using static gauge $X_0 = \kappa
\tau$, the energy density $\mathcal{E} = \sqrt{\lambda} \kappa / 2
\pi$ is uniform along the string so that the string effectively
becomes infinitely long in this limit. Likewise on the gauge side, the
dual single-trace conformal operator of the form $\text{tr}(Z^{J_1}
W^{J_2})$ clearly becomes infinitely long in this limit. If we relax
the trace condition then we are able to consider elementary
excitations on the gauge side given by infinitely long operators of
the form
\begin{equation} \label{magnon operator}
\mathcal{O}_{\mathfrak{p}} = \sum_l e^{i \mathfrak{p} l} (\ldots
ZZZ\underset{\underset{l}\uparrow}WZZZ \ldots).
\end{equation}
The trace condition is equivalent to the requirement that the total
momentum of all excitations should vanish. The single excitation
\eqref{magnon operator}, which violates the momentum condition,
describes a `magnon' of momentum $\mathfrak{p}$ with dispersion
relation \cite{Dorey:2006dq, Beisert:2004hm, Staudacher:2004tk,
Beisert:2005fw, Beisert:2005tm}
\begin{equation*}
E - J_1 = \sqrt{1 + \frac{\lambda}{\pi^2} \sin^2
\frac{\mathfrak{p}}{2}}.
\end{equation*}
At large $\lambda$ this state is described by a classical string
solution on the real line called a `giant magnon' which was
identified in \cite{Hofman:2006xt}. It corresponds to a solitonic
solution of the infinite string embedded in an $\mathbb{R} \times
S^2$ subsector of $AdS_5 \times S^5$.

Subsequently, solitonic solutions of the infinite string moving
through $\mathbb{R} \times S^3$ referred to as `dyonic giant
magnons' were identified in \cite{Dorey:2006dq} and constructed in
\cite{Chen:2006ge}. These solutions carry an extra finite $U(1)$
R-charge $J_2$ and have the following dispersion relation
\begin{equation*}
E - J_1 = \sqrt{J_2^2 + \frac{\lambda}{\pi^2} \sin^2
\frac{\mathfrak{p}}{2}}.
\end{equation*}
They correspond on the gauge side to bound states of $J_2$
magnons given by infinitely long operators of the form
\begin{equation*}
\mathcal{O}_{\mathfrak{p}} = \sum_l e^{i \mathfrak{p} l} (\ldots
ZZZ\underset{\underset{l}\uparrow}W^{J_2}ZZZ \ldots).
\end{equation*}

A general description of such dyonic giant magnons was then
proposed in \cite{Minahan:2006bd} using the language of finite-gap
solutions and spectral curves \cite{KMMZ, Beisert:2004ag,
Beisert:2005bm, Schafer-Nameki:2004ik, Alday:2005gi}, still
restricting attention to the $\mathbb{R} \times S^3$ sector. In this
framework, every solution is characterised by a spectral curve
$\Sigma$ equipped with an Abelian integral $p$ called the
quasi-momentum, such that the pair $(\Sigma, dp)$ encodes the
integrals of motion of the solution. A single dyonic giant magnon can
be described by a condensate cut $\mathcal{B}_1$ from $x_1$ to
$\bar{x}_1$ on the spectral curve \cite{Minahan:2006bd}, whose
presence can be traced down to the existence of a nonvanishing
$a$-period for the differential of the quasi-momentum
\begin{equation*}
\int_a dp \in 2 \pi \mathbb{Z}.
\end{equation*}
The ensuing multivaluedness of the Abelian integral $p(x)$ can
equivalently be described in terms of simple poles of $p(x)$ at
the end points $x_1, \bar{x}_1$ of each condensate cut
$\mathcal{B}_1$. However, this new feature of the quasi-momentum,
which was shown in \cite{Minahan:2006bd} to correctly account for
dyonic giant magnon solutions in the $J_1 \rightarrow \infty$
limit, does not appear in the context of finite-gap integration
simply because the $a$-periods of the differential $dp$ can always
be removed by appropriately normalising $dp$ \cite{KMMZ}. This
apparent dilemma is resolved by noting that the general finite-gap
solution to the equations of a closed string moving through
$\mathbb{R} \times S^3$, constructed in \cite{Paper1, Paper2}, are
valid for all values of the R-charges $J_1$ and $J_2$. It therefore
ought to be possible to obtain dyonic giant magnon solutions as a
special limit of finite-gap solutions when $J_1 \rightarrow \infty$,
and in particular recover the condensate cut representation of
\cite{Minahan:2006bd} in this limit.

In a more recent paper \cite{Okamura:2006zv}, a family of
solutions to the string equations of motion on $\mathbb{R} \times
S^3$ was found by exploiting its connection with the complex
sine-Gordon model via Pohlmeyer's reduction. These solutions
nicely interpolate between folded/circular strings on the one hand
and dyonic giant magnons on the other, the latter being obtained
in the $J_1 \rightarrow \infty$ limit. The analytic form of these
solutions, involving ratios of elliptic $\Theta$-functions and an
overall exponential both of which exhibit a linear $(\sigma,
\tau)$-dependence, is very reminiscent of the general finite-gap
solution for a closed string moving on $\mathbb{R} \times S^3$
constructed in \cite{Paper1, Paper2}.

The aim of this paper is to identify the precise way in which
finite-gap solutions degenerate into dyonic giant magnon solutions
on the real line when taking the $J_1 \rightarrow \infty$ limit.
Inspired by the results of \cite{Okamura:2006zv} we achieve this
by first showing that the (type ($i$)) helical solutions with two
spins of \cite{Okamura:2006zv} are exactly elliptic finite-gap
solutions of \cite{Paper1, Paper2}. The $J_1 \rightarrow \infty$
limit of these solutions is then reinterpreted in the language of
spectral curves. We find that dyonic giant magnons are obtained by
degeneration of the spectral curve $\Sigma$, underlying the
elliptic finite-gap solution, into a singular curve. In
particular, the HM limit can be succinctly summed up as the limit
where the modulus $k$ of the elliptic curve $\Sigma$ goes to
unity. This corresponds to shrinking the real homology period of
$\Sigma$ (e.g. the $b$-period if the cycles are chosen such that
$\hat{\tau} b = b$).

In a general integrable $2$-d field theory there is a well known
connection between solutions to soliton equations on the line and
finite-gap solutions on the circle \cite{Novikov:1984id}. The
former can be described as a limit where the spectral curve of the
latter becomes singular. This phenomenon has been studied in the
case of the KdV and non-linear Schr\"odinger equations as well as
other specific integrable field theories \cite{Belokolos}. The
present paper is the first step towards an investigation of such a
connection in the case of the string equations of motion on
$\mathbb{R} \times S^3$. In the pair of papers \cite{Dressing1,
Dressing2}, a way of constructing a generic configuration of giant
magnons for the equations of motion of the infinite string on
$\mathbb{R} \times S^5$ was presented which makes use of the
dressing method for solitons. The subset of these solutions on
$\mathbb{R} \times S^3$ should be intimately related to singular
limits of the general finite-gap solution constructed in
\cite{Paper1, Paper2} by an extension of the singularisation
procedure, which we identify in the present paper, to higher genus
finite-gap solutions. We hope to come back to this issue soon
\cite{to appear}. Moreover, in the general theory of integrable
$2$-d field theories it is possible to apply this singularisation
procedure only partially to the curve $\Sigma$ (i.e. where only
certain cuts are shrunk to singular points) which generally leads
to more solutions describing solitons scattering on the background
of a finite-gap solution. We do not yet fully understand the
relevance of such partial degenerations in the present context of
string theory on $\mathbb{R} \times S^3$, but hope to come back to
this point in \cite{to appear}.

An investigation of finite-size corrections to giant magnons was
initiated in \cite{Arutyunov:2006gs}. There, a particular
one-soliton solution describing a finite $J_1$ magnon was
considered. Given that giant magnons can be obtained as
degenerations of finite-gap solutions in the $J_1 \rightarrow \infty$
limit, which are valid for arbitrary values of $J_1, J_2$, we 
propose that finite-gap solutions should provide the general
finite-size corrections to giant magnons.

The paper is organised as follows. In section \ref{section: fg} we
recall the construction of the general finite-gap solution to the
equations of a string on $\mathbb{R} \times S^3$ \cite{Paper1,
Paper2}. In section \ref{section: elliptic} we restrict attention
to the elliptic finite-gap solution and explicitly express every
ingredient in the solution in terms of elliptic integrals and
elliptic functions, reducing any integral to the standard elliptic
integrals of the first, second and third kind. As a result, the
general elliptic finite-gap solution is shown to be equivalent to
the (type ($i$)) helical solution of \cite{Okamura:2006zv}. Using
this general two-cut solution, in section \ref{section: singular}
we identify the nature of the HM limit in the finite-gap language.
We also reconcile this picture of dyonic giant magnons as singular
limits of finite-gap solutions with their description in terms of
condensate cuts. Details of the calculations in section
\ref{section: elliptic} are relegated to a series of four
appendices.

\section{Finite-gap string on $\mathbb{R} \times S^3$} \label{section: fg}

The starting point for the method of finite-gap integration is to
rewrite the equations of motion
\begin{equation*}
d \ast j = 0, \quad dj - j \wedge j = 0, \qquad j \in \mathfrak{su}(2)
\end{equation*}
which should be supplemented by the Virasoro constraints $\frac{1}{2}
\text{tr} \, j_{\pm}^2 = - \kappa^2$ (working in conformal static
gauge), as the flatness condition
\begin{equation} \label{flatness}
dJ(x) - J(x) \wedge J(x) = 0,
\end{equation}
for the Lax connection $J(x) = \frac{j - x \ast j}{1 - x^2} \in
\mathfrak{sl}(2,\mathbb{C})$, which depends on the spectral
parameter $x \in \mathbb{C}$. The flatness \eqref{flatness} of the
current $J(x)$ immediately allows one to construct an infinite
number of conserved quantities for the string, which can be neatly
encoded in the spectral curve
\begin{equation*}
\Gamma: \quad \Gamma(x,y) = \text{det} \, (y {\bf 1} -
\Omega(x,\sigma,\tau)) = 0,
\end{equation*}
where the monodromy matrix $\Omega(x,\sigma,\tau) = P
\overleftarrow{\exp} \int_{[c_{\sigma,\tau}]} J(x) \in
SL(2,\mathbb{C})$ is one of the principal objects in integrable
field theories. A non-singular version of this curve can be
obtained, which we hence-forth call $\Sigma$. For generic values
of $x$ the monodromy matrix $\Omega(x,\sigma,\tau)$ has two
distinct eigenvalues, and hence the curve $\Gamma$ (or $\Sigma$)
is a double sheeted ramified cover of $\mathbb{CP}^1$ with
hyperelliptic projection $\hat{\pi} : \Sigma \rightarrow
\mathbb{CP}^1, P \mapsto x$; we define the notation $\{ x^{\pm} \}
= \hat{\pi}^{-1}(x)$ for the set of points above $x \in
\mathbb{CP}^1$. The hyperelliptic curve $\Sigma$ is equipped with
a hyperelliptic holomorphic involution $\hat{\sigma}: \Sigma
\rightarrow \Sigma$ which exchanges the two sheets
$\hat{\sigma}(x^{\pm}) = x^{\mp}$ as well as an anti-holomorphic
involution $\hat{\tau}: \Sigma \rightarrow \Sigma$ which maps both
sheets to themselves by $x \mapsto \bar{x}$ and derives from
reality conditions on $\Omega(x,\sigma,\tau)$ \cite{Paper1}. In
this setup, the dynamical variables are described by a line bundle
\begin{equation*}
L_{\sigma,\tau} \rightarrow \Sigma
\end{equation*}
which encodes the eigenvector $\bm{\psi}(P), P \in \Sigma$ of
$\Omega(x,\sigma,\tau)$. Thus every solution $j$ of the equations of
motion specifies a unique curve $\Sigma$ and line bundle
$L_{\sigma,\tau}$ over this curve; we shall assume that $\Sigma$ has
finite-genus in which case the solution $j$ from which it originates
is called a \textit{finite-gap solution}. Now equation
\eqref{flatness} is the consistency condition of the auxiliary linear
problem
\begin{equation} \label{auxiliary}
(d - J(x)) \bm{\psi} = 0,
\end{equation}
so that \eqref{auxiliary} admits a solution for $\bm{\psi}(P)$
only if $J(x)$ satisfies \eqref{flatness}. The main idea of the
method of finite-gap integration is to note that $\bm{\psi}(P)$ is
uniquely specified by its analytic properties in $P \in \Sigma$
which can be read off from \eqref{auxiliary}, (these properties
define what is called a Baker-Akhiezer vector on $\Sigma$ relative
to the data $\{ \hat{\gamma}, s_{\pm} \}$)
\begin{gather*}
(\psi_1) \geq \hat{\gamma}^{-1} \infty^-, \quad \psi_1(\infty^+) =
1, \quad \text{ and } \quad (\psi_2) \geq \hat{\gamma}^{-1}
\infty^+, \quad \psi_2(\infty^-) = 1,\\
\text{with } \quad \left\{
\begin{split}
&\psi_i(x^{\pm},\sigma,\tau) e^{\mp s_+(x,\sigma,\tau)} = O(1),
\quad \text{as } x \rightarrow 1,\\
&\psi_i(x^{\pm},\sigma,\tau) e^{\mp s_-(x,\sigma,\tau)} = O(1),
\quad \text{as } x \rightarrow -1,
\end{split}
\right. \notag
\end{gather*}
where $s_{\pm}(x,\sigma,\tau) = \frac{i \kappa}{2} \frac{\tau \pm
\sigma}{1 \mp x}$ is obtained from the Virasoro constraints. Thus
the line bundle $L_{\sigma,\tau} \rightarrow \Sigma$ uniquely
specifies an equivalence class of divisors $[\hat{\gamma}]$ of
degree $\text{deg} \; \hat{\gamma} = g+1$. In fact one can show
\cite{Paper2}, taking particular care of the global $SU(2)_R$ degrees
of freedom of the string, that this construction leads to an injective
map
\begin{equation} \label{injective map}
j \mapsto \{ \Sigma, dp, \hat{\gamma} \},
\end{equation}
where the differential $dp$ on $\Sigma$ is such that the pair
$(\Sigma,dp)$ encodes all the moduli of the spectral curve $\Gamma$.
The reconstruction of finite-gap solutions can now be achieved by
constructing the left-inverse of the injective map \eqref{injective
map}. This is done with the help of special functions on the Riemann
surface $\Sigma$ known as Riemann $\theta$-functions. The
reconstruction of the sigma-model field $g \in SU(2)$, out of which $j =
- g^{-1} dg$ is built, requires the concept of the dual Baker-Akhiezer
vector $\bm{\psi}^+$. It is constructed in such a way that it obeys the
following orthogonality relation with the Baker-Akhiezer vector
\begin{equation} \label{dual BA 1}
\bm{\psi}^+(P) \cdot \bm{\psi}(P) = 1, \quad \bm{\psi}^+(\hat{\sigma}
P) \cdot \bm{\psi}(P) = 0.
\end{equation}
Explicitly its components are given as follows
\begin{equation*}
\psi_1^+(P) = \chi(P) \widetilde{\psi}_1^+(P), \quad \psi_2^+(P) =
\frac{\chi(P)}{\chi(\infty^-)} \widetilde{\psi}_2^+(P),
\end{equation*}
where $\chi(P)$ is meromorphic on $\Sigma$ with divisor $(\chi) =
\hat{\gamma} \cdot \hat{\tau} \hat{\gamma} \cdot B^{-1}$ ($B$ being the
divisor of branch points of $\Sigma$) and normalised by
$\chi(\infty^+) = 1$, and $\widetilde{\bm{\psi}}^+(P)$ is a
Baker-Akhiezer vector on $\Sigma$ relative to the data $\{ \hat{\tau}
\hat{\gamma}, - s_{\pm} \}$, which therefore satisfies the reality
condition
\begin{equation} \label{dual BA 2}
\bm{\psi}(\hat{\tau} P)^{\dag} = \widetilde{\bm{\psi}}^+(P).
\end{equation}
It follows using both \eqref{dual BA 1} and \eqref{dual BA 2} that
when $\hat{\tau} P = P$ (or equivalently when $\hat{\pi}(P) \in
\mathbb{R}$) one has
\begin{equation*}
\psi_1(\hat{\sigma} P) = - \frac{A(P)}{\chi(\infty^-)}
\overline{\psi_2(P)}, \quad \psi_2(\hat{\sigma} P) = A(P)
\overline{\psi_1(P)},
\end{equation*}
where $A(P) \equiv \chi(P) \det \left(\bm{\psi}(P),
\bm{\psi}(\hat{\sigma} P) \right)$. Now by definition of $j = -
g^{-1} dg$, the matrix $g^{-1} \in SL(2,\mathbb{C})$ satisfies
$dg^{-1} = j g^{-1}$ and hence since $j = J(0)$ we have
\begin{equation*}
g^{-1} = \frac{1}{\sqrt{\det \Psi(0)}} \Psi(0),
\end{equation*}
up to an $SL(2,\mathbb{C})$ transformation $g \mapsto \tilde{g}_L
\cdot g$ and where
\begin{equation*}
\begin{split}
\Psi(0) = \left( \bm{\psi}(0^+), \bm{\psi}(0^-) \right) &= \left( \begin{array}{cc}
\psi_1(0^+) & \psi_1(0^-)\\ \psi_2(0^+) & \psi_2(0^-) \end{array}
\right),\\
&= \left( \begin{array}{cc}
\psi_1(0^+) & - \frac{1}{\chi(\infty^-)} \overline{\psi_2(0^+)}\\
\psi_2(0^+) & \overline{\psi_1(0^+)} \end{array} \right) \text{diag}
\left( 1, A(0^+) \right),\\
&= \tilde{g}_R \left( \begin{array}{cc}
\psi_1(0^+) & - \chi(\infty^-)^{-\frac{1}{2}} \overline{\psi_2(0^+)}\\
\chi(\infty^-)^{-\frac{1}{2}} \psi_2(0^+) & \overline{\psi_1(0^+)}
\end{array} \right) \tilde{g}_L,
\end{split}
\end{equation*}
with $\tilde{g}_R = \text{diag}\left( 1,
\chi(\infty^-)^{\frac{1}{2}} \right)$ and $\tilde{g}_L =
\text{diag}\left( 1, \chi(\infty^-)^{- \frac{1}{2}} A(0^+)
\right)$. Thus finally, after a residual $SL(2,\mathbb{C})_R \times
SL(2,\mathbb{C})_L$ transformation (c.f. \cite{Paper1} p43) the
reconstructed matrix $g^{-1}$ lives in $SU(2)$,
\begin{equation} \label{matrix g}
g = \left( \begin{array}{cc} Z_1 & Z_2\\ - \bar{Z_2} & \bar{Z_1}
\end{array} \right),
\end{equation}
where
\begin{equation} \label{fg solution}
Z_1 = C \widetilde{\psi}^+_1(0^+), \quad Z_2 =
\frac{C}{\chi(\infty^-)^{\frac{1}{2}}} \widetilde{\psi}^+_2(0^+)
\end{equation}
and $C \in \mathbb{R}$ is a normalisation constant chosen such that
$|Z_1|^2 + |Z_2|^2 = 1$.

It is clear from the above construction that the (dual) Baker-Akhiezer
vector does not depend on the choice of canonical homology basis for
$H_1(\Sigma,\mathbb{Z})$ and so neither does the general finite-gap solution
\eqref{fg solution}. However, when explicitly constructing the
solution in terms of $\theta$-functions on $\Sigma$ as we will do
below a particular choice of canonical homology basis for
$H_1(\Sigma,\mathbb{Z})$ is required. Different choices of $a$- and
$b$-cycles, connected to one another by an $\text{Sp}(2g,\mathbb{Z})$
transformation, offer alternative but equivalent parametrisations of
one and the same finite-gap solution. Furthermore, one is also free to
make a different choice of branch cuts when representing the Riemann
surface $\Sigma$ as a two-sheeted ramified cover of $\mathbb{CP}^1$
since the cuts are not an intrinsic property of $\Sigma$. These remark
will be essential later when we come to study the Hofman-Maldacena
limit of the finite-gap solutions.


The Baker-Akhiezer vector $\widetilde{\bm{\psi}}^+$ can be
reconstructed explicitly using Riemann $\theta$-functions, in
particular we find for the components of
$\widetilde{\bm{\psi}}^+(0^+)$
\begin{subequations} \label{reconstruction formula for psi^+}
\begin{equation} \label{reconstruction formula for psi^+_1}
\widetilde{\psi}^+_1(0^+) = h_-(0^+) \frac{\theta \big(\bm{D}; \Pi \big) \theta
\big( 2 \pi \int^{0^+}_{\infty^+} \bm{\omega} - \int_{\bm{b}}
d\mathcal{Q} - \bm{D}; \Pi \big)}{\theta \big(\int_{\bm{b}}
d\mathcal{Q} + \bm{D}; \Pi \big) \theta \big( 2 \pi
\int^{0^+}_{\infty^+} \bm{\omega} - \bm{D}; \Pi \big)} \exp \left( +
\frac{i}{2} \int_{\infty^-}^{\infty^+} d\mathcal{Q} - \frac{i}{2}
\int_{0^-}^{0^+} d\mathcal{Q} \right),
\end{equation}
\begin{equation} \label{reconstruction formula for psi^+_2}
\widetilde{\psi}^+_2(0^+) = h_+(0^+) \frac{\theta \big(\bm{D}; \Pi \big) \theta
\big( 2 \pi \int^{0^+}_{\infty^-} \bm{\omega} - \int_{\bm{b}}
d\mathcal{Q} - \bm{D}; \Pi \big)}{\theta \big(\int_{\bm{b}}
d\mathcal{Q} + \bm{D}; \Pi \big) \theta \big( 2 \pi
\int^{0^+}_{\infty^-} \bm{\omega} - \bm{D}; \Pi \big)} \exp \left( -
\frac{i}{2} \int_{\infty^-}^{\infty^+} d\mathcal{Q} - \frac{i}{2}
\int_{0^-}^{0^+} d\mathcal{Q} \right).
\end{equation}
\end{subequations}
Here $h_{\pm}(P)$ are meromorphic functions on $\Sigma$ defined as
follows
\begin{equation*}
(h_{\pm}) \geq \infty^{\pm} (\hat{\gamma}^+)^{-1}, \quad
 h_{\pm}(\infty^{\mp}) = 1,
\end{equation*}
and the vector $\bm{D} \equiv \bm{\mathcal{A}}(\hat{\gamma}^+) -
\bm{\mathcal{A}}(\infty^-) + \bm{\mathcal{K}} \in \mathbb{C}^g$ is
almost\footnote{`almost' refers to the fact that
$\theta(\bm{\mathcal{A}}(P) - \bm{D})$ shouldn't vanish
identically.} arbitrary, where $\bm{\mathcal{A}}(P) = 2 \pi
\int_{\infty^+}^P \bm{\omega}$ is the Abel map\footnote{The
integration contour from $\infty^+$ to $P$ in $\bm{\mathcal{A}}(P)
= 2 \pi \int_{\infty^+}^P \bm{\omega}$ is taken to lie within the
normal form $\Sigma_{\text{cut}}$ of $\Sigma$ defined by cutting
the Riemann surface $\Sigma$ along the cycles $a,b$. This
corresponds to a choice of branch for the multi-valued Abelian
integral $\int^P \bm{\omega}$.} and $\bm{\mathcal{K}}$ is the
vector of Riemann constants which can be related in a simple way
to the canonical class $Z$ (the divisor class of any meromorphic
differential), namely \cite{G&H}
\begin{equation} \label{K relation to Z}
2 \bm{\mathcal{K}} = - \bm{\mathcal{A}}(Z).
\end{equation}
The components $\{ \omega_i \}_{i=1}^g$ of the vector
$\bm{\omega}$ are the basis holomorphic differentials on $\Sigma$
defined by their $a$-periods
\begin{equation*}
\int_{a_i} \omega_j = \delta_{ij},
\end{equation*}
and in terms of which the period matrix $\Pi$ can be defined
\begin{equation*}
\Pi_{ij} = \int_{b_i} \omega_j.
\end{equation*}
A special role is played in the formulae \eqref{reconstruction formula
for psi^+} by the differential $d\mathcal{Q} = \frac{1}{2 \pi}(\sigma
dp + \tau dq)$ where $dp, dq$ are the normalised\footnote{An
Abelian differential $\Omega$ of the second or third kind is said to be
\textit{normalised} if its $a$-periods are all set to zero,
$\int_{a_i} \Omega = 0, i = 1, \ldots, g$.} differentials of the
quasi-momentum and quasi-energy. Finally,
\begin{equation} \label{theta def}
\theta(\bm{z} ; \Pi) \equiv \sum_{\bm{m} \in \mathbb{Z}^g} \exp
\left\{ i \langle \bm{m} , \bm{z} \rangle + \pi i \langle \Pi \bm{m},
\bm{m} \rangle \right\}, \quad \text{for } \bm{z} \in \mathbb{C}^g
\end{equation}
is the Riemann $\theta$-function associated with the Riemann surface
$\Sigma$.


By uniqueness of the (dual) Baker-Akhiezer vector the reconstruction
formulae \eqref{reconstruction formula for psi^+} give the same
solution for any choice of basis of $H_1(\Sigma,\mathbb{Z})$, but the
reality conditions on the various ingredients in \eqref{reconstruction
formula for psi^+} are dependent on this choice of basis. This is
because we may choose a basis for which the $a$-cycles say are imaginary
$\hat{\tau} a_i = - a_i$ and the $b$-cycles are real $\hat{\tau} b_i =
b_i$, but we may just as well choose a basis for which the reverse is
true, $\hat{\tau} a_i = a_i$ and $\hat{\tau} b_i = - b_i$. This does
not mean that the solution exhibits different reality properties for
different choices of $a$- and $b$-cycles, but only that the parameters
of the solution may have different reality properties for different
choices of cycles; the resulting matrix $g$ in \eqref{matrix g} is
always $SU(2)$ valued. Equation \eqref{K relation to Z} allows us to
reformulate the reality conditions on
$\bm{\mathcal{A}}(\hat{\gamma}^+)$ obtained in \cite{Paper1} (with
respect to the canonical homology basis of that paper), as reality
conditions on the vector $\bm{D}$ since
\begin{equation} \label{real D}
\begin{split}
2 \text{Im} \, \bm{\mathcal{A}}(\hat{\gamma}^+) &= \bm{\mathcal{A}}
\left( Z \cdot (\infty^-)^2 \cdot (\infty^+)^2 \right)\\
&= \bm{\mathcal{A}}(Z) + 2 \bm{\mathcal{A}}(\infty^-)\\
&= 2 \bm{\mathcal{A}}(\infty^-) - 2 \bm{\mathcal{K}},\\
\Rightarrow \quad \text{Im} \, \bm{D} &= 0,
\end{split}
\end{equation}
where we have used the fact that $\overline{\bm{\mathcal{K}}} = -
\bm{\mathcal{K}}$ and $\overline{\bm{\mathcal{A}}(\infty^-)} = -
\bm{\mathcal{A}}(\infty^-)$ hold as equalities on the Jacobian
$J(\Sigma)$, which themselves follow from $\overline{\hat{\tau}^{\ast}
\bm{\omega}} = - \bm{\omega}$.

Let us also specify an alternative definition to \eqref{K relation to Z} of
the vector of Riemann constants $\bm{\mathcal{K}}$ which will come in
handy later. One can show with our conventions that the components of
$\bm{\mathcal{K}}$ are given by the following formulae
\begin{equation} \label{K def}
\mathcal{K}_k = 2 \pi \left[ \frac{1 + \Pi_{k k}}{2} - \sum_{j=1,
j\neq k}^g \int_{a_j}\left( \int^P_{\infty^+} \omega_k \right)
\omega_j \right].
\end{equation}

\section{Elliptic (two-gap) case} \label{section: elliptic}

When the underlying curve $\Sigma$ of the previous section is elliptic
(genus $1$), everything within the expressions \eqref{fg
solution}, \eqref{reconstruction formula for psi^+} for the general
finite-gap solution can be explicitly computed in terms of elliptic
functions. The aim of this section is to obtain the most general
elliptic finite-gap solution. To avoid cluttering this section with
lengthy calculations we shall refer to a series of appendices for the
details.

\begin{figure}
\centering \psfrag{X1}{\footnotesize{$x_1$}}
\psfrag{X1b}{\footnotesize{$\bar{x}_1$}}
\psfrag{X2}{\footnotesize{$x_2$}}
\psfrag{X2b}{\footnotesize{$\bar{x}_2$}}
\psfrag{a}{\footnotesize{$a$}} \psfrag{b}{\footnotesize{$b$}}
\psfrag{x}{\tiny{$x$}}
\includegraphics[width=50mm]{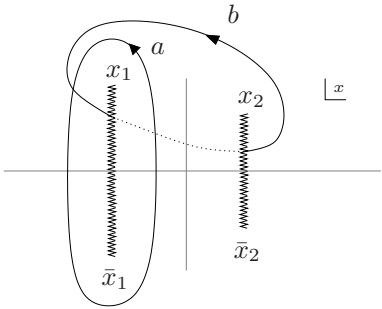}
\caption{$a$- and $b$-periods in $x$-plane.} \label{a b periods}
\end{figure}

So consider the most general (real\footnote{The reality condition on
the curve must be imposed, which simply requires the set of branch
points $\{ x_1, x_2, \bar{x}_1, \bar{x}_2 \}$ to be invariant
under conjugation $x \mapsto \bar{x}$, see \cite{Paper1} for
details.}) elliptic curve given algebraically by
\begin{equation} \label{curve}
y^2 = (x - x_1)(x - \bar{x}_1)(x - x_2)(x - \bar{x}_2),
\end{equation}
which can be represented as a $2$-sheeted Riemann surface with $2$
cuts. The $a$- and $b$-periods of the curve are
chosen\footnote{Note that for the $b$-period we are using a
slightly different convention to that in \cite{Paper1}. In
\cite{Paper1} the $b$-period would join $x_2$ to $\bar{x}_1$
\begin{center}
\begin{tabular}{c} \psfrag{X1}{\tiny{$x_1$}}
\psfrag{X1b}{\tiny{$\bar{x}_1$}}
\psfrag{X2}{\tiny{$x_2$}}
\psfrag{X2b}{\tiny{$\bar{x}_2$}}
\psfrag{a}{\tiny{$a$}}
\psfrag{b}{\tiny{$b$}}
\psfrag{x}{\tiny{$x$}}
\includegraphics[width=35mm]{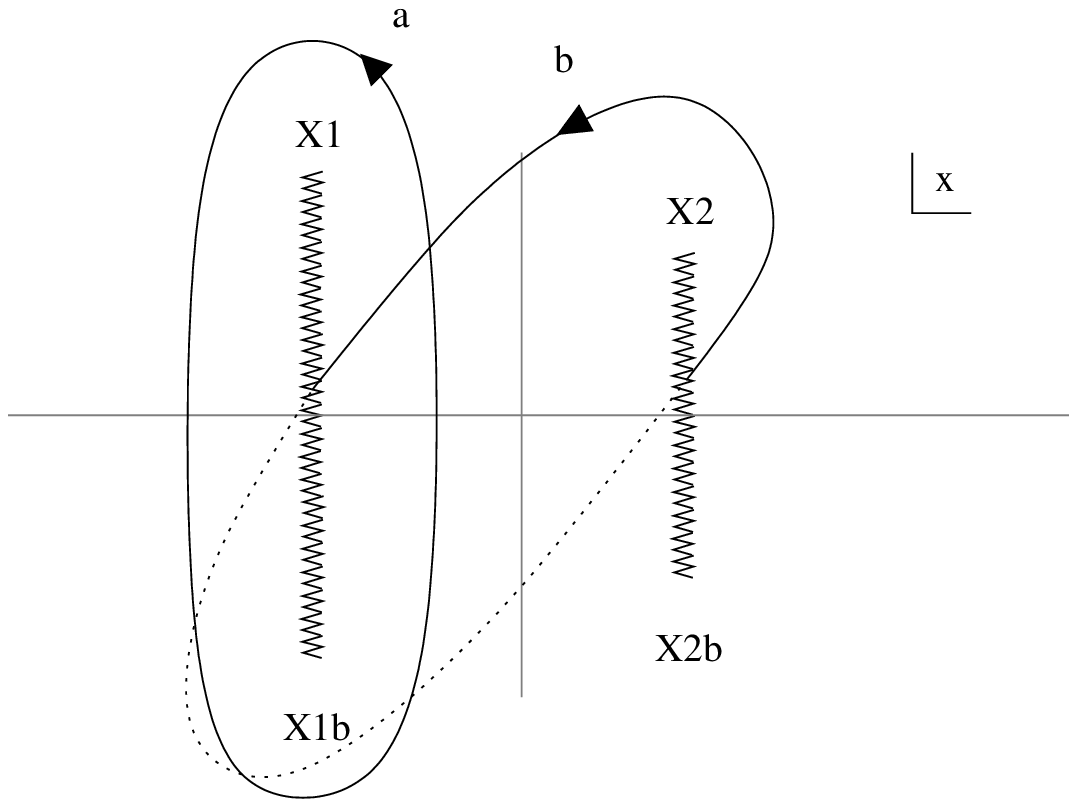}\end{tabular}
\end{center}
This change of convention is convenient in the elliptic case under
consideration here because in this case (and only this case) the
period matrix $\Pi$ (which is a single complex number $\tau$ in
this case) turns out to be purely imaginary. } as in Figure \ref{a
b periods}. On this curve we can define a unique normalised (vanishing
$a$-period) holomorphic differential
\begin{equation*}
\omega = \nu \Big/ \int_a \nu,
\end{equation*}
where $\nu = dx/y$ is a holomorphic differential on \eqref{curve}. The
period matrix in the elliptic case is just a single complex number
which we compute in appendix \ref{section: Elliptic setup} to be,
\begin{equation*}
\Pi = \int_b \omega = \int_b \nu \Big/ \int_a \nu = \frac{i K'}{K} \equiv \tau,
\end{equation*}
where $K(k)$ and $K'(k) = K(k')$ are the elliptic integrals of the
first kind depending on the elliptic modulus $k$ of the curve and its
dual modulus $k'$ defined as,
\begin{equation*}
k' = \left| \frac{x_1 - x_2}{x_1 - \bar{x}_2} \right|, \quad k =
\sqrt{1 - (k')^2}.
\end{equation*}

The Riemann $\theta$-function reduces in the elliptic case to
the Jacobi $\vartheta$-function
\begin{equation*}
\vartheta_3(z;\tau) = \sum_{n = -\infty}^{\infty} \exp (\pi i n^2
\tau + 2 \pi i n z) = \theta(2 \pi z; \tau).
\end{equation*}
It is useful to define three other Jacobi $\vartheta$-functions as
translations of $\vartheta_3(z;\tau)$ by half-periods, namely
\begin{equation*}
\begin{split}
\vartheta_0(z;\tau) &= \vartheta_3 \left(z + \frac{1}{2}; \tau\right),\\
\vartheta_1(z;\tau) &= \exp\left(\pi i \frac{\tau}{4} + \pi i
\left(z + \frac{1}{2}\right)\right) \vartheta_3 \left(z + \frac{\tau
+ 1}{2}; \tau\right),\\
\vartheta_2(z;\tau) &= \exp\left(\pi i \frac{\tau}{4} + \pi i z
\right) \vartheta_3 \left(z + \frac{\tau}{2}; \tau\right).
\end{split}
\end{equation*}
We also introduce the Jacobi $\Theta$-functions which
depend directly on the elliptic modulus $k$ of the elliptic curve
\begin{equation*}
\Theta_{\mu}(z; k) \equiv \vartheta_{\mu}\left( \frac{z}{2K}; \tau
= \frac{iK'}{K} \right), \qquad \mu = 0,\ldots,3.
\end{equation*}
When there is no ambiguity as to what the elliptic modulus $k$ is we
will omit it from the arguments and simply write $\Theta_{\mu}(z) =
\Theta_{\mu}(z; k)$.

In appendix \ref{section: Theta functions} we compute explicitly in
the elliptic case the various ingredients appearing inside the
$\theta$-functions of \eqref{reconstruction formula for psi^+}. We
show for instance that the real parts of $\int_{\infty^{\pm}}^{0^+}
\omega$ are $\frac{1}{2}$ and $0$ respectively so that we can write
\begin{equation} \label{rho def}
\int_{\infty^+}^{0^+} \omega = \frac{1}{2} + i \rho_+, \qquad
\int_{\infty^-}^{0^+} \omega = \frac{\tau}{2} + i \rho_-.
\end{equation}
Here $\rho_{\pm} \in \mathbb{R}$ are real numbers function of the
moduli of the elliptic curve. The reason for the shift by
$\frac{\tau}{2}$ in the second of these integrals (which could be
absorbed into the arbitrary constant $\rho_-$) will become clear
later. At this stage we can already simplify the reconstruction
formulae \eqref{fg solution} a bit
\begin{subequations} \label{reconstruction formula for Z}
\begin{equation} \label{reconstruction formula for Z_1}
Z_1 = C h_-(0^+) \frac{\vartheta_0 \big( X_0; \tau \big) \vartheta_3
\big( X - i \rho_+; \tau \big)}{\vartheta_3 \big( X_0 - i \rho_+;
\tau \big) \vartheta_0 \big( X; \tau \big)} \; \exp \left( -i
\int_{\infty^+}^{0^+} d\mathcal{Q} \right),
\end{equation}
\begin{equation} \label{reconstruction formula for Z_2}
Z_2 = C \frac{h_+(0^+)}{\chi(\infty^-)^{\frac{1}{2}}}
\frac{\vartheta_0 \big( X_0; \tau \big) \vartheta_1 \big( X - i
\rho_-; \tau \big)}{\vartheta_1 \big( X_0 - i \rho_-; \tau \big)
\vartheta_0 \big( X; \tau \big)} \; \exp \left( -i
\int_{\infty^-}^{0^+} d\mathcal{Q}  + \frac{i}{2} \int_b d\mathcal{Q}
\right),
\end{equation}
\end{subequations}
where we have defined $X \equiv \frac{1}{2 \pi} \int_b
d\mathcal{Q} + X_0$ and $X_0 \equiv \frac{1}{2 \pi} D - \frac{1}{2}
\in \mathbb{R}$, or if we define
the notation $\tilde{A} = 2 K A$
then
\begin{subequations} \label{reconstruction formula for Z3}
\begin{equation} \label{reconstruction formula for Z3_1}
Z_1 = C h_-(0^+) \frac{\Theta_0 \big( \tilde{X}_0 \big) \Theta_3 \big(
\tilde{X} - i \tilde{\rho}_+ \big)}{\Theta_3 \big( \tilde{X}_0 - i
\tilde{\rho}_+ \big) \Theta_0 \big( \tilde{X} \big)} \; \exp
\left( -i \int_{\infty^+}^{0^+} d\mathcal{Q} \right),
\end{equation}
\begin{equation} \label{reconstruction formula for Z3_2}
Z_2 = C \frac{h_+(0^+)}{\chi(\infty^-)^{\frac{1}{2}}} \frac{\Theta_0
\big( \tilde{X}_0 \big) \Theta_1 \big( \tilde{X} - i \tilde{\rho}_-
\big)}{\Theta_1 \big( \tilde{X}_0 - i \tilde{\rho}_- \big) \Theta_0
\big( \tilde{X} \big)} \; \exp \left( -i \int_{\infty^-}^{0^+}
d\mathcal{Q}  + \frac{i}{2} \int_b d\mathcal{Q} \right).
\end{equation}
\end{subequations}
Expressions for the variables $\rho_{\pm}$ in terms of the
moduli of the curve can also be obtained. In appendix \ref{section:
Theta functions} we derive a closed form expression for the
integrals in \eqref{rho def}, namely
\begin{equation} \label{int rel final}
\int_{\infty^{\pm}}^{0^+} \omega = \frac{i F(\varphi_{\mp}, k')}{2
K},
\end{equation}
where,
\begin{equation} \label{phi final 2}
\tan \frac{\varphi_{\pm}}{2} =
\frac{\left( \sqrt{\bar{x}_2} \pm \sqrt{x_1} \right) \left(
\sqrt{\bar{x}_1} + \sqrt{x_2} \right) }{|x_1 - \bar{x}_2|}.
\end{equation}
From these equations we conclude that the real variables
$\rho_{\pm}$ depend not only on the magnitude $k' = |h|$ of $h$
but also on another real parameter, related to the phase of $h$.
This is just what we require to exhibit the fact that
$\rho_{\pm}$ do contain a free parameter of the solution. However,
because we are dealing with a $2$-cut finite-gap solution there
cannot be more than $3$ independent conserved quantities
\cite{Paper1, Paper2}: the two global charges $R$ and $L$ or
equivalently $J_1 = (L-R)/2$ and $J_2 = (L+R)/2$, and the single
internal charge (recall \cite{Paper1, Paper2} that in the general
case with genus $g$ there are $g$ internal charges).

Gathering together equations \eqref{rho def} and \eqref{int rel
final} we have $i \tilde{\rho}_- + i K' = i F(\varphi_+,k')$ from which
one can deduce that
\begin{equation*}
i \rho_- = 0 \Leftrightarrow \tan \frac{\varphi_+}{2} = 1 \Leftrightarrow
\arg \left( \frac{x_1}{\bar{x}_2} \right) \in \pi \mathbb{Z}.
\end{equation*}
An example which meets this condition is when the set of branch points
$\{ x_1, x_2, \bar{x}_1, \bar{x}_2\}$ of the curve has the extra
symmetry $x \rightarrow -x$. However, with $\bar{x}_2 = - x_1$ we find
$\tan \frac{\varphi_-}{2} = -i$. Then, putting together the
equations \eqref{rho def} and \eqref{int rel final} we have $K + i
\tilde{\rho}_+ = i F(\varphi_-,k')$, which yields
\begin{equation*}
\rho_+ = 0.
\end{equation*}
It is clear then that the Frolov-Tseytlin solution (for which the
curve has the extra symmetry $x \rightarrow -x$) corresponds to the
case $\rho_{\pm} = 0$. In this limit the space of solutions breaks up
into two distinct non-singular sectors:
\begin{figure}[t]
\centering \psfrag{x1}{\tiny $x_1$} \psfrag{x1b}{\tiny
$\bar{x}_1$} \psfrag{x2}{\tiny $x_2$}
\begin{tabular}{ccc}
\includegraphics[width=40mm]{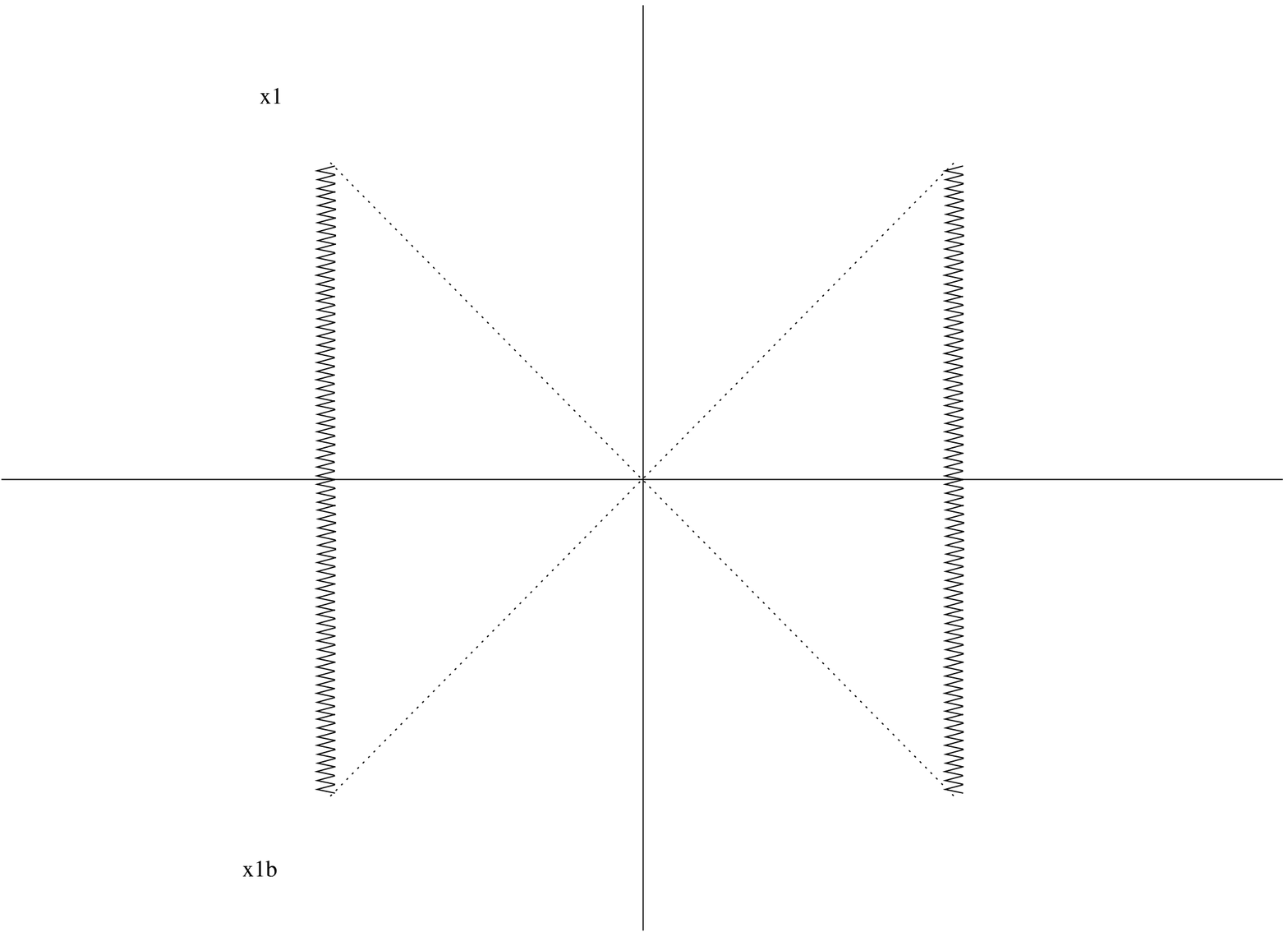} & &
\includegraphics[width=40mm]{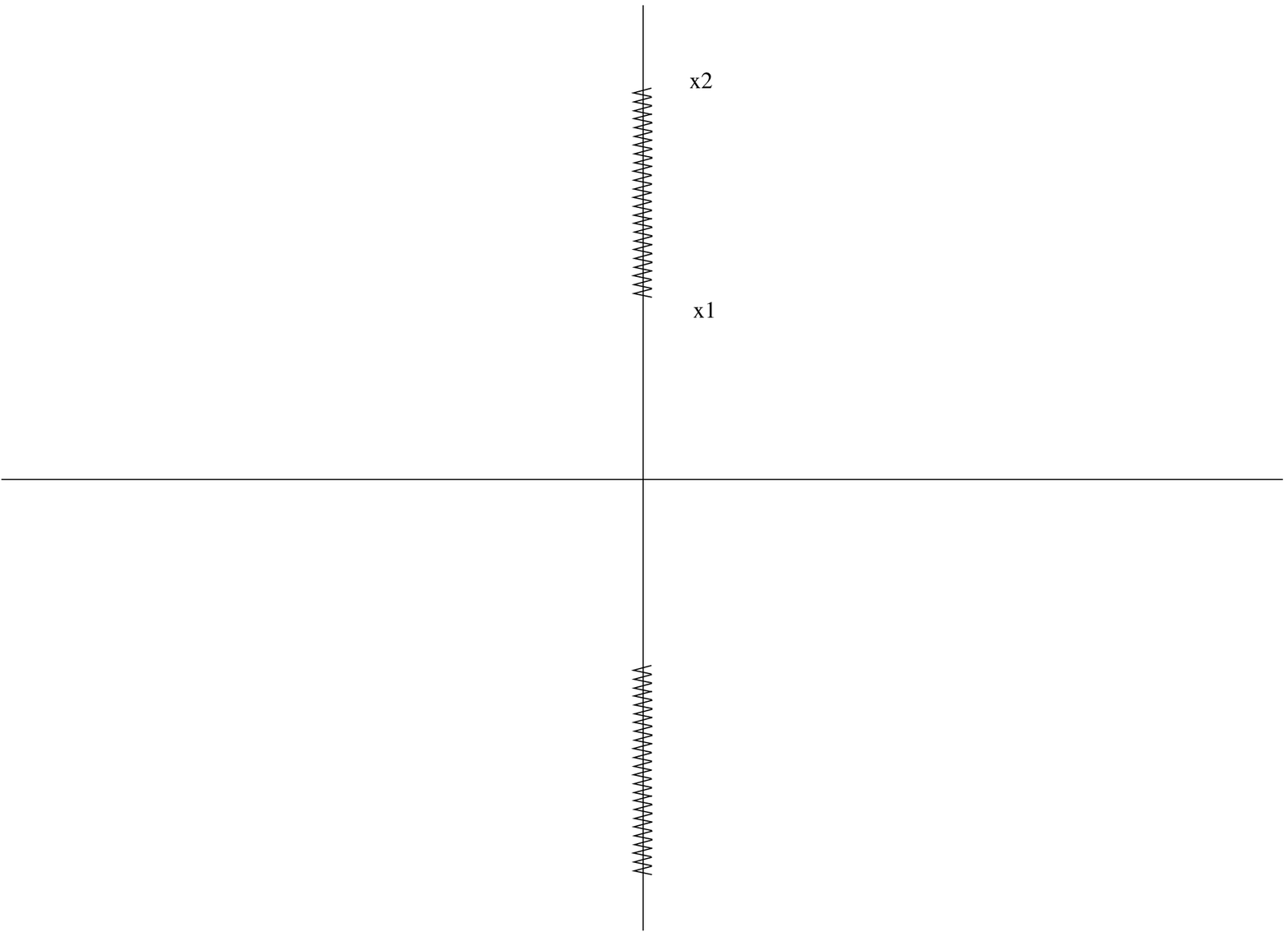}\\
$(i)$ & & $(ii)$
\end{tabular}
\caption{two distinct possible limits of elliptic finite-gap solutions under
Frolov-Tseytlin limit $\rho_{\pm} \rightarrow 0$.}
\label{FT limit}
\end{figure}
\begin{itemize}
\item[$(i)$] $x_1 = - \bar{x}_2$, in which case the curve
$\Sigma$ is parametrised by a single complex number $x_1 \in
\mathbb{C}$, and corresponds to the ``double contour''
configuration of Bethe roots on the gauge theory side (see Figure
\ref{FT limit} $(i)$).
\item[$(ii)$] $x_1 = - \bar{x}_1, x_2 = - \bar{x}_2$, in which
case the curve $\Sigma$ is parametrised by two real numbers $i
x_1, i x_2 \in \mathbb{R}$, and corresponds to the ``imaginary
root'' configuration of Bethe roots (see Figure \ref{FT limit}
$(ii)$).
\end{itemize}
To go from one sector to the other without violating the reality
condition or the symmetry $x \rightarrow -x$ one has to go through
the common singular limit $k \rightarrow 1$ of both sectors. So
the two sectors $(i),(ii)$ are effectively disconnected regions of
the parameter space for non-singular $\Sigma$ in the
Frolov-Tseytlin limit $\rho_{\pm} \rightarrow 0$.

Next we obtain the $b$-periods for the differential
$d\mathcal{Q}$. Using the Riemann bilinear identities for the
differentials $dp$ and $\omega$ one finds
\begin{equation*}
\int_b dp = - 2 \pi i \sum \text{res} \; p \omega,
\end{equation*}
where the Abelian integral $p(P) = \int^P dp$ has simple poles at
$x = \pm 1$ of the form\footnote{In \cite{Paper1, Paper2} the
overall sign was different and we had $p(x) \underset{x
\rightarrow \pm 1}{\sim} - \frac{\pi \kappa}{x \mp 1}$. However,
this difference of sign simply comes down to the choice of the
physical sheet.}
\begin{equation} \label{p asymp}
p(x) \sim \frac{\pi \kappa}{x \mp 1} \qquad \text{as} \quad x
\rightarrow \pm 1.
\end{equation}
And so, using expression \eqref{omega def} for the holomorphic
differential $\omega$ we obtain
\begin{equation} \label{dp b period}
\frac{1}{2 \pi} \int_b dp = \frac{\pi \kappa |x_1 - \bar{x}_2|}{2
K} \left( \frac{1}{y_+} + \frac{1}{y_-} \right),
\end{equation}
where $y_{\pm} = \left. y(x)\right|_{x = \pm 1} > 0$ by the choice
of branch for the function $y$ (see appendix \ref{section:
Elliptic setup}). Likewise, for the Abelian integral $q(P) =
\int^P dq$ whose simple poles at $x = \pm 1$ are of the form
\begin{equation} \label{q asymp}
q(x) \sim \pm \frac{\pi \kappa}{x \mp 1} \qquad \text{as} \quad x
\rightarrow \pm 1,
\end{equation}
one obtains
\begin{equation} \label{dq b period}
\frac{1}{2 \pi} \int_b dq = \frac{\pi \kappa |x_1 - \bar{x}_2|}{2
K} \left( \frac{1}{y_+} - \frac{1}{y_-} \right).
\end{equation}
Now equations \eqref{dp b period} and \eqref{dq b period} together
imply
\begin{equation*}
\left( \frac{1}{2 \pi} \int_b dp \right)^2 - \left( \frac{1}{2 \pi}
\int_b dq \right)^2 = \frac{\pi^2 \kappa^2 |x_1 - \bar{x}_2|^2}{K^2 y_+ y_-},
\end{equation*}
so that if we define rescaled coordinates
\begin{equation*}
(x,t) = (\kappa' \sigma,\kappa' \tau) \qquad \text{where} \quad
\kappa' \equiv \kappa \frac{|x_1 - \bar{x}_2|}{\sqrt{y_+ y_-}},
\end{equation*}
then
\begin{equation*}
\begin{split}
X - X_0 = \frac{1}{2 \pi} \int_b d\mathcal{Q} &= \left( \frac{1}{4
\pi^2} \int_b dp \right) \sigma + \left( \frac{1}{4 \pi^2} \int_b
dq \right) \tau,\\ &= \frac{1}{2} \left( \sqrt{\frac{y_-}{y_+}} +
\sqrt{\frac{y_+}{y_-}} \right) \frac{x}{2 K} + \frac{1}{2} \left(
\sqrt{\frac{y_-}{y_+}} - \sqrt{\frac{y_+}{y_-}} \right) \frac{t}{2
K}.\\
&= \frac{1}{2 K} \frac{x - v t}{\sqrt{1 - v^2}},
\end{split}
\end{equation*}
where
\begin{equation*}
v \equiv \frac{y_+ - y_-}{y_+ + y_-}.
\end{equation*}
Note that $|v| < 1$ since $y_{\pm} >0$. So the scaled variable
$\tilde{X}$ appearing in \eqref{reconstruction formula for Z3}
satisfies
\begin{equation*}
\tilde{X} = \tilde{X_0} + \frac{x - v t}{\sqrt{1 - v^2}}.
\end{equation*}
Let us also define a boosted time coordinate
\begin{equation*}
\tilde{T} = \frac{1}{2} \left( \sqrt{\frac{y_-}{y_+}} +
\sqrt{\frac{y_+}{y_-}} \right) t + \frac{1}{2} \left(
\sqrt{\frac{y_-}{y_+}} - \sqrt{\frac{y_+}{y_-}} \right) x = \frac{t -
v x}{\sqrt{1 - v^2}}.
\end{equation*}

In appendix \ref{section: Exponentials} we express the exponents
in \eqref{reconstruction formula for psi^+} in terms of elliptic
integrals. The details of this computation are not important and
so we simply state the result here. The general elliptic
finite-gap solutions \eqref{reconstruction formula for Z3} now
takes the following form
\begin{subequations} \label{reconstruction formula for Z4}
\begin{equation} \label{reconstruction formula for Z4_1}
Z_1 = C h_-(0^+) \frac{\Theta_0 \big( \tilde{X}_0 \big) \Theta_3
\big( \tilde{X} - i \tilde{\rho}_+ \big)}{\Theta_3 \big(
\tilde{X}_0 - i \tilde{\rho}_+ \big) \Theta_0 \big( \tilde{X}
\big)} \; \exp \left( Z_2(i \tilde{\rho}_+,k) \left( \tilde{X} -
\tilde{X}_0 \right) + i v_+ \tilde{T} \right),
\end{equation}
\begin{equation} \label{reconstruction formula for Z4_2}
Z_2 = C \frac{h_+(0^+)}{\chi(\infty^-)^{\frac{1}{2}}}
\frac{\Theta_0 \big( \tilde{X}_0 \big) \Theta_1 \big( \tilde{X} -
i \tilde{\rho}_- \big)}{\Theta_1 \big( \tilde{X}_0 - i
\tilde{\rho}_- \big) \Theta_0 \big( \tilde{X} \big)} \; \exp
\left( Z_0(i \tilde{\rho}_-,k) \left( \tilde{X} - \tilde{X}_0
\right) + i v_- \tilde{T} \right),
\end{equation}
\end{subequations}
where we have defined
\begin{equation*}
v_{\pm} = \frac{y(0) \pm 1}{|x_1 - \bar{x}_2|},
\end{equation*}
which satisfy $v_-^2 - v_+^2 = \text{dn}^2 (i \tilde{\rho}_-,k) +
(k')^2 \text{sc}^2 (i \tilde{\rho}_+,k)$.

Finally, all that remains to be determined are the normalisation
constants in \eqref{reconstruction formula for Z4} which are obtained
in appendix \ref{section: Normalisation},
\begin{equation} \label{prefactors}
\begin{split}
h_-(0^+) &= \frac{\vartheta_3 \left( X_0 - i \rho_+ \right)}{\vartheta_2
\left( i \rho_+ \right)} = \frac{\Theta_3 \left(
\tilde{X}_0 - i \tilde{\rho}_+ \right)}{\Theta_2
\left( i \tilde{\rho}_+ \right)}, \\
\frac{h_+(0^+)}{\chi(\infty^-)^{\frac{1}{2}}} &= \frac{\vartheta_1 \left(
X_0 - i \rho_- \right)}{\vartheta_0 \left( i \rho_- \right)} =
\frac{\Theta_1 \left( \tilde{X}_0 - i \tilde{\rho}_- \right)}{\Theta_0
\left( i \tilde{\rho}_- \right)}.
\end{split}
\end{equation}
Plugging these constants back into the general elliptic finite-gap
solutions \eqref{reconstruction formula for Z4} yields
\begin{subequations} \label{reconstruction formula for Z5}
\begin{equation} \label{reconstruction formula for Z5_1}
Z_1 = C \frac{\Theta_3 \big( \tilde{X} - i \tilde{\rho}_+
\big)}{\Theta_2 \big( i \tilde{\rho}_+ \big) \Theta_0 \big(
\tilde{X} \big)} \; \exp \left( Z_2(i \tilde{\rho}_+,k) \tilde{X}
+ i v_+ \tilde{T} + i \varphi^0_1 \right),
\end{equation}
\begin{equation} \label{reconstruction formula for Z5_2}
Z_2 = C \frac{\Theta_1 \big( \tilde{X} - i \tilde{\rho}_-
\big)}{\Theta_0 \big( i \tilde{\rho}_- \big) \Theta_0 \big(
\tilde{X} \big)} \; \exp \left( Z_0(i \tilde{\rho}_-,k) \tilde{X}
+ i v_- \tilde{T} + i \varphi^0_2 \right),
\end{equation}
\end{subequations}
where we have introduced global phases $\varphi^0_i, i=1,2$ into which
we have absorbed the terms in the exponentials involving
$\tilde{X}_0$. Notice that this solution corresponds exactly (up to a
trivial interchange of coordinates $Z_1 \leftrightarrow Z_2$) to the
type $(i)$ helical string with two spins of \cite{Okamura:2006zv} when
$X_0 = \varphi^0_1 = \varphi^0_2 = 0$. However \eqref{reconstruction
formula for Z5} does contain the extra `initial value' degree of
freedom compared to the type $(i)$ helical string which corresponds to
the initial value of the internal degree of freedom. The initial
values $\varphi^0_i, i=1,2$ of the global $SU(2)_R \times SU(2)_L$
degrees of freedom have also been trivially included.

\subsection{Global charges}

In this section we discuss the global conserved charges $L,R$
corresponding to the Casimirs of the global $SU(2)_R \times
SU(2)_L$ symmetry as well as the space-time energy $E$ of the
string corresponding to translation invariance in the target time
coordinate $X_0$. The moduli of the elliptic curve $\Sigma$ which
encodes the conserved quantities of the solution can be succinctly
described in terms of a special differential \cite{Paper1}
\begin{equation} \label{alpha def}
\alpha \equiv \frac{\sqrt{\lambda}}{4 \pi} \left( x + \frac{1}{x}
\right) dp,
\end{equation}
where $dp$ is the quasi-momentum invoked previously. In particular,
the global charges $L,R$ are both expressed in terms of the residues
of $\alpha$ at $0^+,\infty^+$ respectively\footnote{The overall sign
here is different from that in \cite{Paper1,Paper2} for the same
reason that the sign in \eqref{p asymp} was different.},
\begin{equation} \label{global charges L,R}
\frac{L}{2} = - \text{res}_{0^+} \alpha, \quad \frac{R}{2} =
\text{res}_{\infty^+} \alpha.
\end{equation}
Let us define from $L,R$ the usual linear combinations $J_1,J_2$ of
conserved charges that are relevant for comparison with the gauge
theory, namely
\begin{equation} \label{global charges J_1,J_2}
\begin{split}
J_1 &= \frac{L + R}{2} = - \text{res}_{0^+} \alpha +
\text{res}_{\infty^+} \alpha,\\
J_2 &= \frac{L - R}{2} = - \text{res}_{0^+} \alpha -
\text{res}_{\infty^+} \alpha.
\end{split}
\end{equation}
In section \ref{section: HM limit} we will want to consider the
Hofman-Maldacena limits of these charges. In this limit we will
see that the two cuts of the elliptic curve merge together to
leave behind a pair of complex conjugate singular points in the
$x$-plane. Thus it will be useful to consider the situation where
$x = 0$ lies outside the region in between the two cuts. The
charge $J_2$ can be broken down into a sum of contributions from
each pair of branch points of $\Sigma$ for we have
\begin{equation} \label{J_2 decomposition}
J_2 = \frac{1}{2 \pi i} \int_{a_1} \alpha + \frac{1}{2 \pi i}
\int_{a_2} \alpha = \frac{1}{2 \pi i} \int_{b_1} \alpha + \frac{1}{2
\pi i}\int_{b_2} \alpha = 2 \text{Re} \left[ \frac{1}{2 \pi i}
\int_{b_1} \alpha \right],
\end{equation}
where the cycles $a_i,b_i, i = 1,2$ are depicted in Figure
\ref{a,b cycles for J_2}.
\begin{figure}[h] \centering
\psfrag{a1}{\textcolor{green}{\tiny $a_1$}}
\psfrag{a2}{\textcolor{green}{\tiny $a_2$}}
\psfrag{b}{\textcolor{green}{\tiny $b_1$}}
\psfrag{b2}{\textcolor{green}{\tiny $b_2$}}
\psfrag{equiv}{\textcolor{blue}{$\equiv$}}
\psfrag{J_2}{\textcolor{blue}{$\rightsquigarrow J_2$}}
\psfrag{0+}{\tiny $0^+$} \psfrag{0-}{\tiny $0^-$}
\psfrag{i+}{\tiny $\infty^+$} \psfrag{i-}{\tiny $\infty^-$}
\begin{tabular}{c}
\includegraphics[height=34mm]{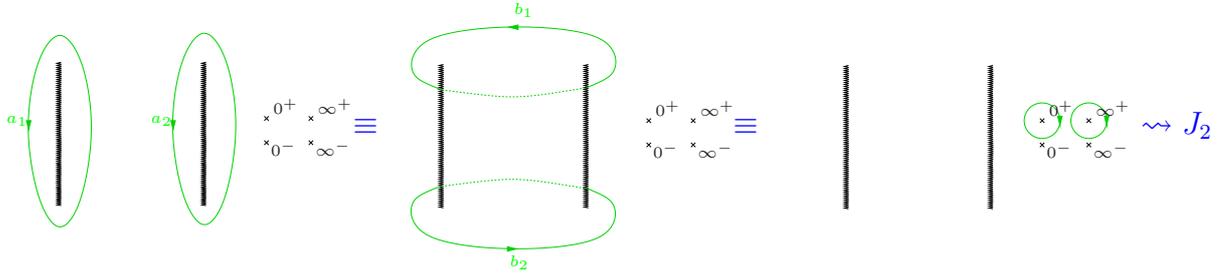}\\
\end{tabular}
\caption{Contributions to global charge $J_2$ from pairs of
branch points.} \label{a,b cycles for J_2}
\end{figure}
The total space-time energy $E$ of the string which has a simple form
in static gauge $X_0 = \kappa \tau$,
\begin{equation} \label{space-time energy}
E \equiv \frac{\sqrt{\lambda}}{2 \pi} \int_0^{2 \pi} d\sigma
\partial_{\tau} X_0 = \kappa \sqrt{\lambda}.
\end{equation}
To obtain a similar expression to \eqref{J_2 decomposition} but
for the other charge $J_2$ we consider the differential
\begin{equation} \label{tilde alpha def}
\tilde{\alpha} \equiv \frac{\sqrt{\lambda}}{4 \pi} \left(x -
\frac{1}{x}\right)dp = \frac{\sqrt{\lambda}}{4 \pi}
\frac{(x-1)(x+1)}{x} dp.
\end{equation}
It has simple poles at $x = \pm1$, $x = 0$ and $x = \infty$ on the top
sheet by virtue of the fact that $dp$ has only double poles at $x =
\pm 1$ of the form
\begin{equation*}
dp(x) = d\left( \frac{\pi \kappa}{x \mp 1} \right) + O\left( (x \mp
1)^0 \right), \quad \text{as} \; x \rightarrow \pm 1.
\end{equation*}
Thus we can write
\begin{equation*}
0 = \sum_{I = 1}^2 \frac{1}{2 \pi i} \int_{a_I} \tilde{\alpha} +
\sum_{x \in \{ \pm 1, 0 , \infty\}} \text{res}_x \tilde{\alpha},
\end{equation*}
which can be simplified using \eqref{tilde alpha def}, the
definitions of $E$ and $J_1$, and the fact that the sum of the
$a_I$-periods of $\tilde{\alpha}$ is equal to the sum of its
$b_I$-periods, yielding
\begin{equation} \label{E - J_1 decomposition}
E - J_1 = \sum_{I = 1}^2 \frac{1}{2 \pi i} \int_{b_I} \tilde{\alpha} =
2 \text{Re} \left[ \frac{1}{2 \pi i} \int_{b_1} \tilde{\alpha}
\right].
\end{equation}
This is to be compared with the expression for $J_2$ in
\eqref{J_2 decomposition}.

\subsection{Periodicity}

All finite-gap solutions are quasi-periodic since the dynamical
divisor $\hat{\gamma}(x,t)$ moves linearly
\begin{equation} \label{linear motion}
\mathcal{A}(\hat{\gamma}(x,t)) =
\mathcal{A}(\hat{\gamma}(x_0,t_0)) + \left( \frac{1}{2 \pi} \int_b
dp \right) \frac{x - x_0}{\kappa'} + \left( \frac{1}{2 \pi} \int_b
dq \right) \frac{t - t_0}{\kappa'} \in J(\Sigma),
\end{equation}
on the Jacobian $J(\Sigma)$ which is a compact complex torus
\cite{Paper1}. Since the string is closed we require the two-cut solution
\eqref{reconstruction formula for Z5} to be real $\sigma$-periodic and
so we must impose the condition
\begin{equation} \label{periodicity condition 1}
\frac{1}{2 \pi} \int_b dp \equiv n \in \mathbb{Z}.
\end{equation}
Equation \eqref{dp b period} then implies
\begin{equation*}
\frac{\pi \kappa |x_1 - \bar{x}_2|}{2 K} \left( \frac{1}{y_+} +
\frac{1}{y_-} \right) = n,
\end{equation*}
which can equivalently be rewritten as
\begin{equation} \label{periodicity1}
\frac{2 K \sqrt{1 - v^2}}{\kappa'} = \frac{2 \pi}{n}.
\end{equation}
It is clear from the periodicity property $\Theta_{\mu}(z + 2 K) =
\Theta_{\mu}(z)$ of $\Theta$-functions that the fundamental period
in the $x$ variable of the $\Theta$-function part of the formula
\eqref{reconstruction formula for Z5} is
\begin{equation} \label{single-hop period}
T_x \equiv 2 K \sqrt{1 - v^2} = \frac{2 \pi \kappa'}{n}.
\end{equation}
Thus we can break up the full closed string interval $x \in [-\pi
\kappa', \pi \kappa' )$ (i.e. $\sigma \in [-\pi, \pi)$) into $n$
equal intervals, referred to as `hops' in \cite{Okamura:2006zv},
corresponding to regions of periodicity of the internal degrees of
freedom. In the next section we shall restrict attention to the
following `single-hop' region
\begin{equation} \label{single hop}
-\frac{1}{2} T_x \leq x < \frac{1}{2} T_x.
\end{equation}
A `single-hop' corresponds in the algebro-geometric language to a
single traverse of the Jacobian $J(\Sigma)$ by the dynamical
divisor $\hat{\gamma}(x,t)$ as is clear from \eqref{linear
motion}, or put another way, a single traverse of a homology
$a$-cycle by the divisor $\hat{\gamma}(x,t)$ on $\Sigma$.

Besides the condition \eqref{periodicity condition 1} which comes
from periodicity requirements on the internal degrees of freedom,
another periodicity condition comes from considering the global
$SU(2)_R \times SU(2)_L$ degrees of freedom in the exponentials of
\eqref{reconstruction formula for Z5}, leading to
\begin{equation} \label{periodicity condition 2}
\frac{1}{2 \pi} \int^{0^+}_{\infty^{\pm}} dp \equiv - N_{\pm} \in
\mathbb{Z}.
\end{equation}
This is the statement that after a full traverse of the string
interval $\sigma \in [0, 2\pi)$ the arguments of the exponentials
should have changed by integer multiples of $2\pi$. Restricting
attention to the `single-hop' region we must require that the
arguments of the exponentials only change by integer multiples of
$\frac{2\pi}{n}$. In other words, combining \eqref{periodicity
condition 2} with equations \eqref{int dQ infty-} and \eqref{int
dQ infty+} yields the final periodicity conditions
\begin{equation} \label{periodicity2}
\begin{split}
&2 K \left[ -i Z_0(i\tilde{\rho}_-,k) - v \cdot v_- \right] - \pi
= \frac{2 \pi N_-}{n},\\ &2 K \left[ -i Z_2(i\tilde{\rho}_+,k) - v
\cdot v_+ \right] = \frac{2 \pi N_+}{n},
\end{split}
\end{equation}
which correspond to the changes in the exponentials over a
`single-hop'.

\section{Singular curve} \label{section: singular}

\subsection{Giant magnon limit} \label{section: HM limit}

In \cite{Okamura:2006zv} the Hofman-Maldacena (HM) limit was
argued in the case of the elliptic solution \eqref{reconstruction
formula for Z5} to correspond to taking the elliptic moduli $k$ to
unity, while at the same time scaling the world-sheet coordinates
by letting $\kappa \rightarrow \infty$. However, the limit $\kappa
\rightarrow \infty$ is actually forced upon us when we take the
limit $k \rightarrow 1$, as can be seen from equation
\eqref{periodicity1}. It follows that the HM limit is simply
\begin{equation*}
\text{HM limit}: \quad k \rightarrow 1.
\end{equation*}
But taking the elliptic modulus $k$ to unity is equivalent to
taking the complementary elliptic modulus $k' = \sqrt{1 - k^2}$ to
zero, which from its definition in \eqref{Mobius} means that in
this limit the branch points $x_1$ and $x_2$ merge. This in turn
is equivalent to the pinching of a particular $b$-cycle of the
curve. However, keeping the solution real as we take this limit
requires that the conjugate branch points $\bar{x}_1$ and
$\bar{x}_2$ also merge, so that one should also pinch another
particular $b$-cycle.
\begin{figure}[h]
\centering \psfrag{a}{\textcolor{red}{\tiny $a$}}
\psfrag{b}{\textcolor{green}{\tiny $b$}}
\begin{tabular}{ccccc}
\includegraphics[height=30mm]{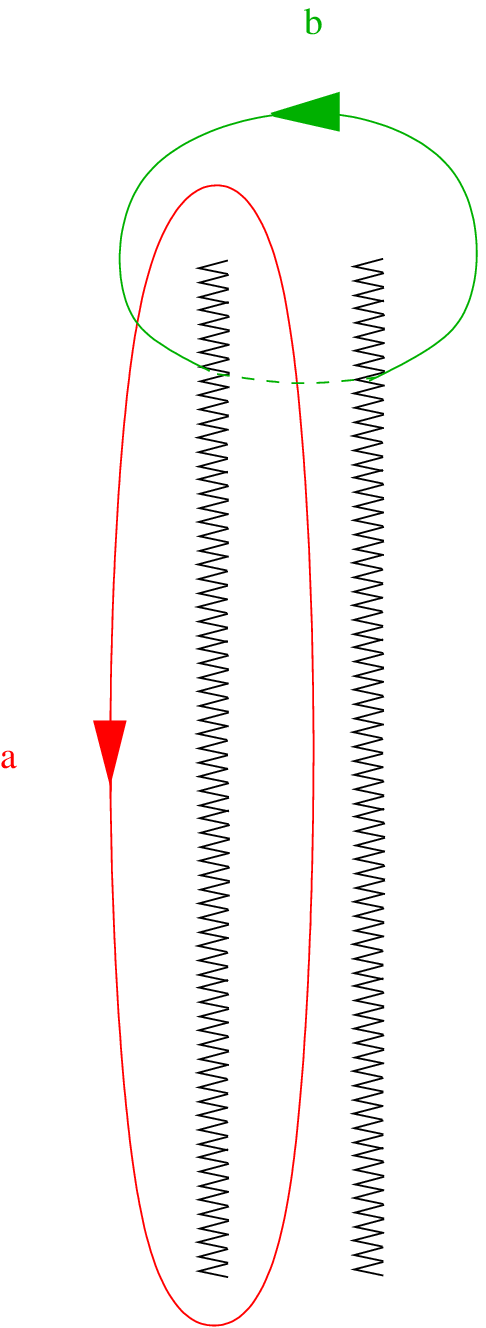} & $\qquad$ &
\includegraphics[height=30mm]{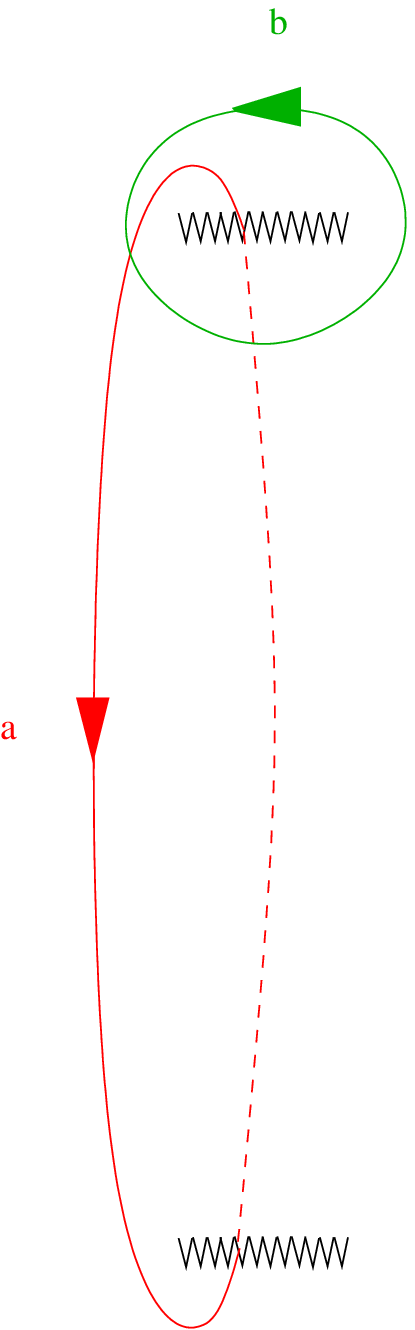} & $\qquad$ &
\includegraphics[height=30mm]{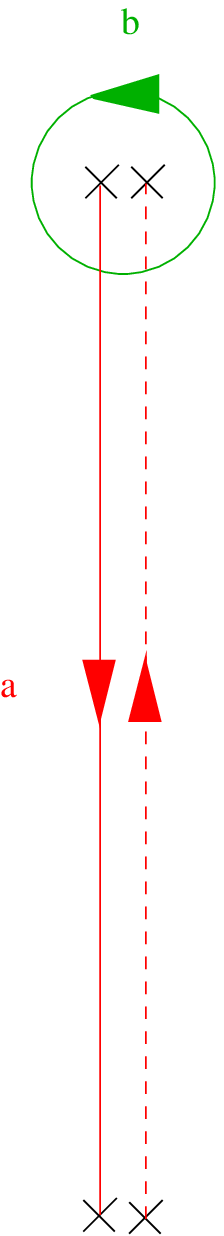}\\
$(a)$ & & $(b)$ & & $(c)$
\end{tabular}
\caption{$(a)$ Neighbouring branch cuts on $\Sigma$. $(b)$ The
same elliptic curve with the branch cuts chosen differently. $(c)$
The singular limit $\Sigma_{\text{sing}}$ of $\Sigma$.} \label{b
degeneration}
\end{figure}
To understand what happens to the curve in this degeneration consider
the situation when $x_1$ is very close to $x_2$, $|x_1 - x_2| \ll 1$
(see Figure \ref{b degeneration} $(a)$). As the branch points merge in
pairs ($x_1$ with $x_2$ and $\bar{x}_1$ with $\bar{x}_2$), the branch
cuts (which can be freely chosen to connect pairs of branch points
which are merging, see Figure \ref{b degeneration} $(b)$) disappear
and leave behind singular points at the now coalescing branch points
$x_1 = x_2$ and $\bar{x}_1 = \bar{x}_2$. In the limit, a homology
$a$-cycle joins the singular points $x_1$ and $\bar{x}_1$ on the top
sheet and goes back from $\bar{x}_1$ to $x_1$ on the bottom sheet,
whereas a homology $b$-cycle circles the singular point $x_1$ on the
top sheet (see Figure \ref{b degeneration} $(c)$).

Since the period $T_x$ of a `single-hop' defined in
\eqref{single-hop period} blows up in the singular curve limit $k
\rightarrow 1$, this means that as the elliptic curve $\Sigma$
degenerates to a singular curve the restriction of the
corresponding periodic two-gap solution \eqref{reconstruction
formula for Z5} to the `single-hop' in \eqref{single hop} turns
into a soliton solution on the real line $x \in \mathbb{R}$.
Indeed, the $k \rightarrow 1$ limit of the two-gap solution
\eqref{reconstruction formula for Z5} looks like
\begin{equation} \label{giant magnon}
Z_1 = \frac{\cos(\tilde{\rho}_-)}{\cosh\big(\tilde{X}\big)} e^{i
v_+ \tilde{T} + i \varphi^0_1}, \quad Z_2 = \frac{
\sinh\big(\tilde{X} - i
\tilde{\rho}_-\big)}{\cosh\big(\tilde{X}\big)} e^{i
\tan(\tilde{\rho}_-) \tilde{X} + i v_- \tilde{T} + i \varphi^0_2},
\end{equation}
where $v_-^2 - v_+^2 = \text{dn}^2(i \tilde{\rho}_-,1) = 1 +
\tan^2 \tilde{\rho}_-$. Using the periodicity condition
\eqref{periodicity2} it follows that (keeping the ratio
$\frac{N_-}{n}$ finite but $\frac{N_+}{n} \sim K$ in the limit $k
\rightarrow 1$)
\begin{equation*}
v = \frac{\tan \tilde{\rho}_-}{v_-}, \quad v_+ = \tan \alpha,
\end{equation*}
for some $\alpha \in \mathbb{R}$, so that (after scaling the
coordinates $(x,t) \rightarrow \left( \cos (\alpha) \, x, \cos
(\alpha) \, t \right)$ by a finite quantity) equation \eqref{giant
magnon} yields exactly the dyonic giant magnon solution (again
after the interchange of coordinates $Z_1 \leftrightarrow Z_2$)
\begin{equation*}
\begin{split}
Z_1 &= \frac{1}{\sqrt{1 + \tilde{k}^2}} \frac{1}{\cosh\big(\cos
(\alpha) \, \tilde{X} \big)} e^{i \sin (\alpha) \, \tilde{T} + i
\varphi^0_1}, \\
Z_2 &= \frac{1}{\sqrt{1 + \tilde{k}^2}} \left[ \tanh\big(\cos
(\alpha) \, \tilde{X}\big) - i \tilde{k} \right] e^{i t + i
\varphi^0_2}.
\end{split}
\end{equation*}
with $\tilde{k} \equiv \tan \tilde{\rho}_-$.

The global charges of this dyonic giant magnon solution are
obtained as the $k \rightarrow 1$ limits of the global charges
\eqref{global charges J_1,J_2} of the elliptic solution. If we defined
the charges $\mathcal{J}_1, \mathcal{J}_2$ and energy $\mathcal{E}$
attributed to a `single-hop' by
\begin{equation*}
\mathcal{J}_i = \frac{J_i}{n}, \quad i = 1,2, \qquad \mathcal{E} =
\frac{E}{n} = \kappa \frac{\sqrt{\lambda}}{n},
\end{equation*}
then we may write down the following linear combinations which remain
finite as $k \rightarrow 1$
\begin{equation} \label{E-J,Q}
\begin{split}
\mathcal{E} - \mathcal{J}_1 &= \frac{\sqrt{\lambda}}{4 \pi}
\left| \left( x_1 - \frac{1}{x_1} \right) - \left( \bar{x}_1 -
\frac{1}{\bar{x}_1} \right) \right|,\\
\mathcal{J}_2 &= \frac{\sqrt{\lambda}}{4 \pi} \left| \left( x_1
+ \frac{1}{x_1} \right) - \left( \bar{x}_1 + \frac{1}{\bar{x}_1}
\right) \right|,
\end{split}
\end{equation}
which follow from \eqref{E - J_1 decomposition} and \eqref{J_2
decomposition} respectively, using the expression for $dp$ (on the top
sheet) in the singular limit
\begin{equation*}
dp(x) = \frac{\pi \kappa dx}{(x - x_1)(x - \bar{x}_1)} \left[
\frac{|1 - x_1|^2}{(x - 1)^2} + \frac{|1 + x_1|^2}{(x + 1)^2}
\right].
\end{equation*}
We also define the giant magnon momentum as
\begin{equation} \label{magnon mom}
\mathfrak{p} \equiv \pi - 2 \tilde{\rho}_- = 2 \pi
\int_{\infty^+}^{0^+} \nu \Big/ \int_b \nu = -i \ln \left(
\frac{x_1}{\bar{x}_1} \right),
\end{equation}
where the last two equalities follow from the singular limit $k
\rightarrow 1$ of the second equation in \eqref{rho def} and using
$K'(1) = \frac{\pi}{2}$. The dyonic giant magnon dispersion
relation follows immediately from \eqref{E-J,Q} and \eqref{magnon
mom}, namely
\begin{equation*}
\mathcal{E} - \mathcal{J}_1 = \sqrt{\mathcal{J}_2^2
+ \frac{\lambda}{\pi^2} \sin^2 \frac{\mathfrak{p}}{2}}.
\end{equation*}

\subsection{Connection with condensate cuts}

In this section we mention the connection between the above
construction of giant magnons as singular limits of finite-gap
solutions with the construction of giant magnons in terms of
curves with `condensate cuts' \cite{Minahan:2006bd}.

\begin{figure}
\centering \psfrag{a}{\textcolor{red}{\tiny $a$}}
\psfrag{b}{\textcolor{green}{\tiny $b$}}
\begin{tabular}{ccccc}
\includegraphics[width=30mm]{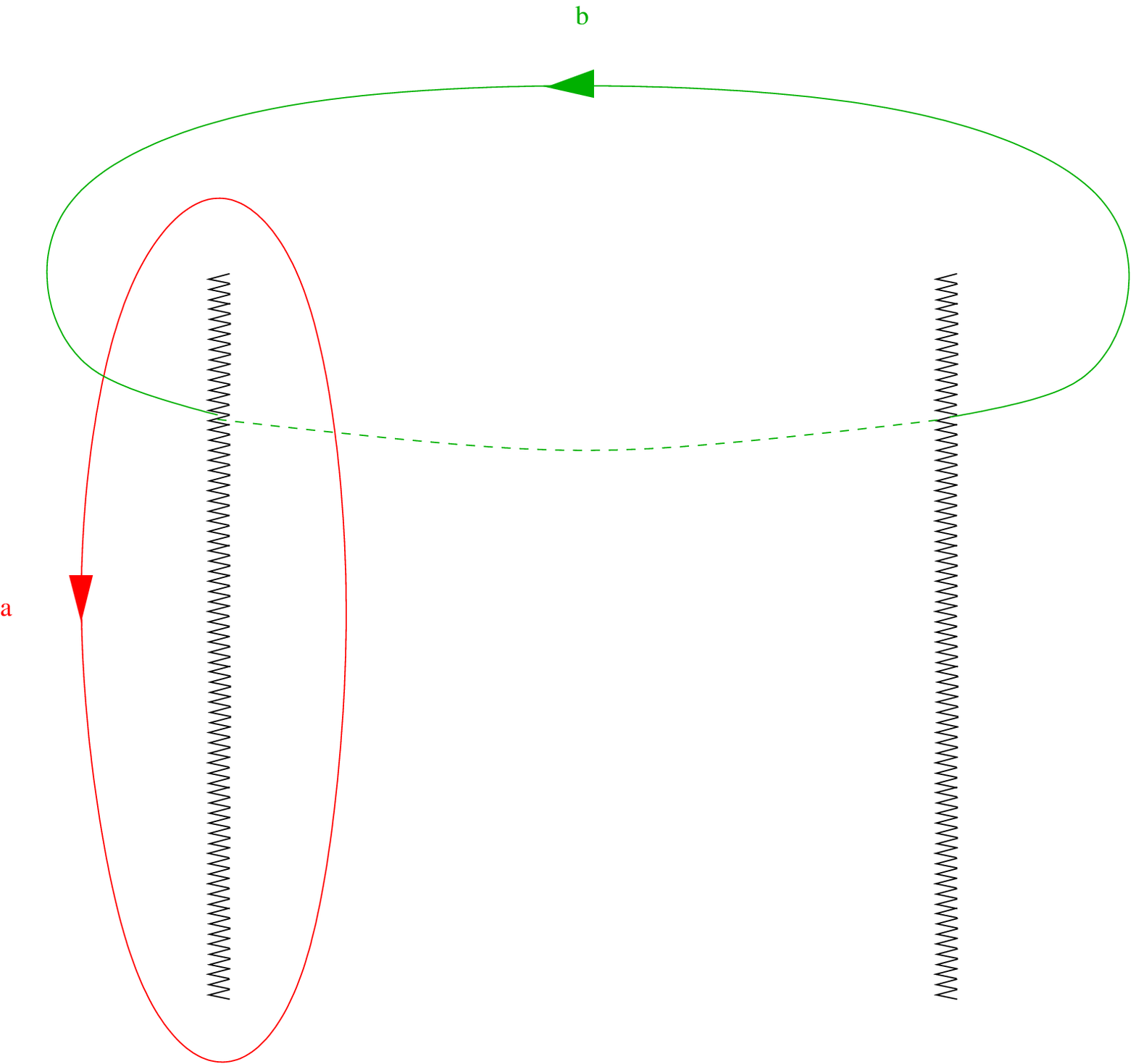} & $\quad$ &
\includegraphics[width=30mm]{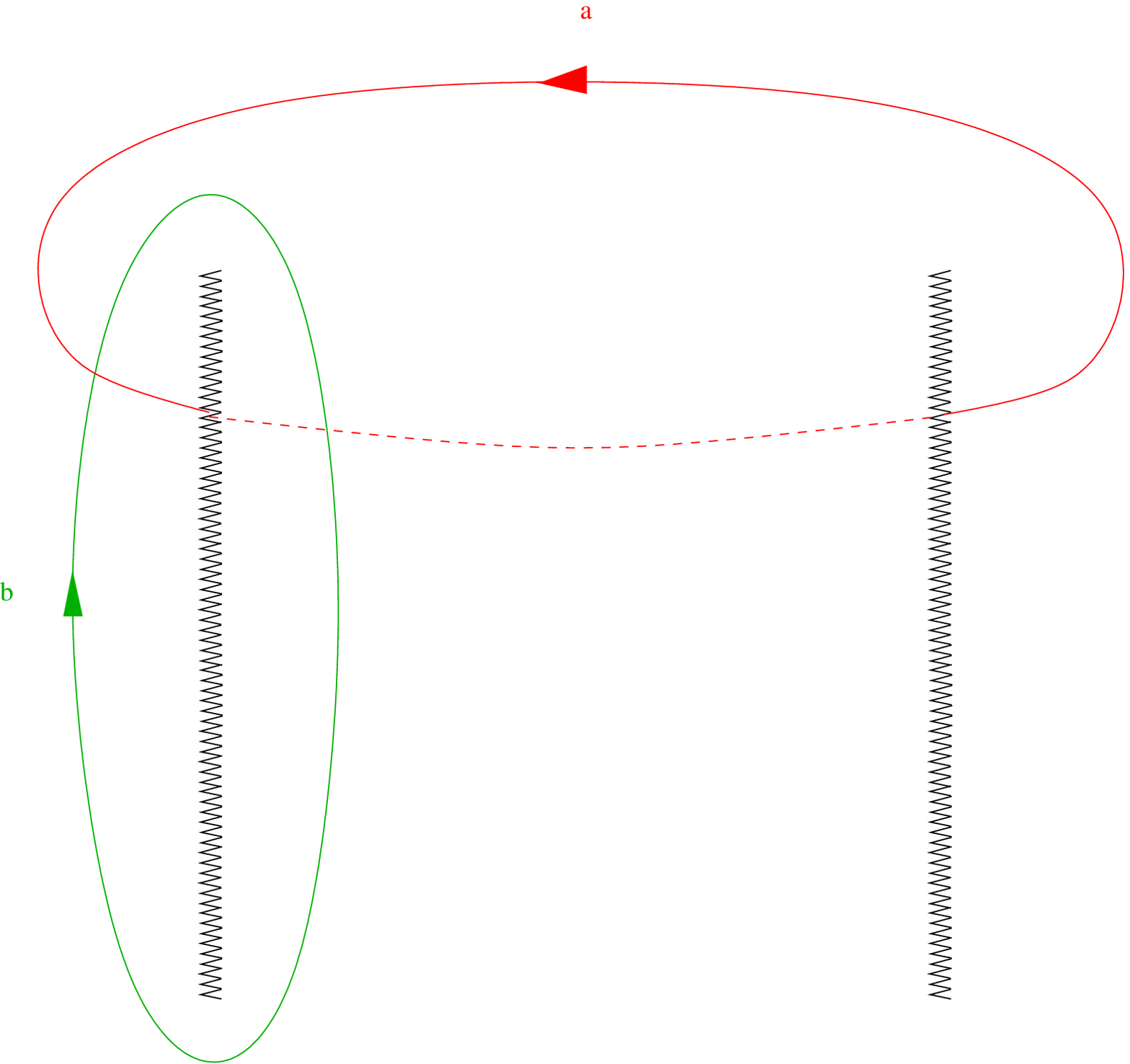} & $\quad$ &
\includegraphics[width=27mm]{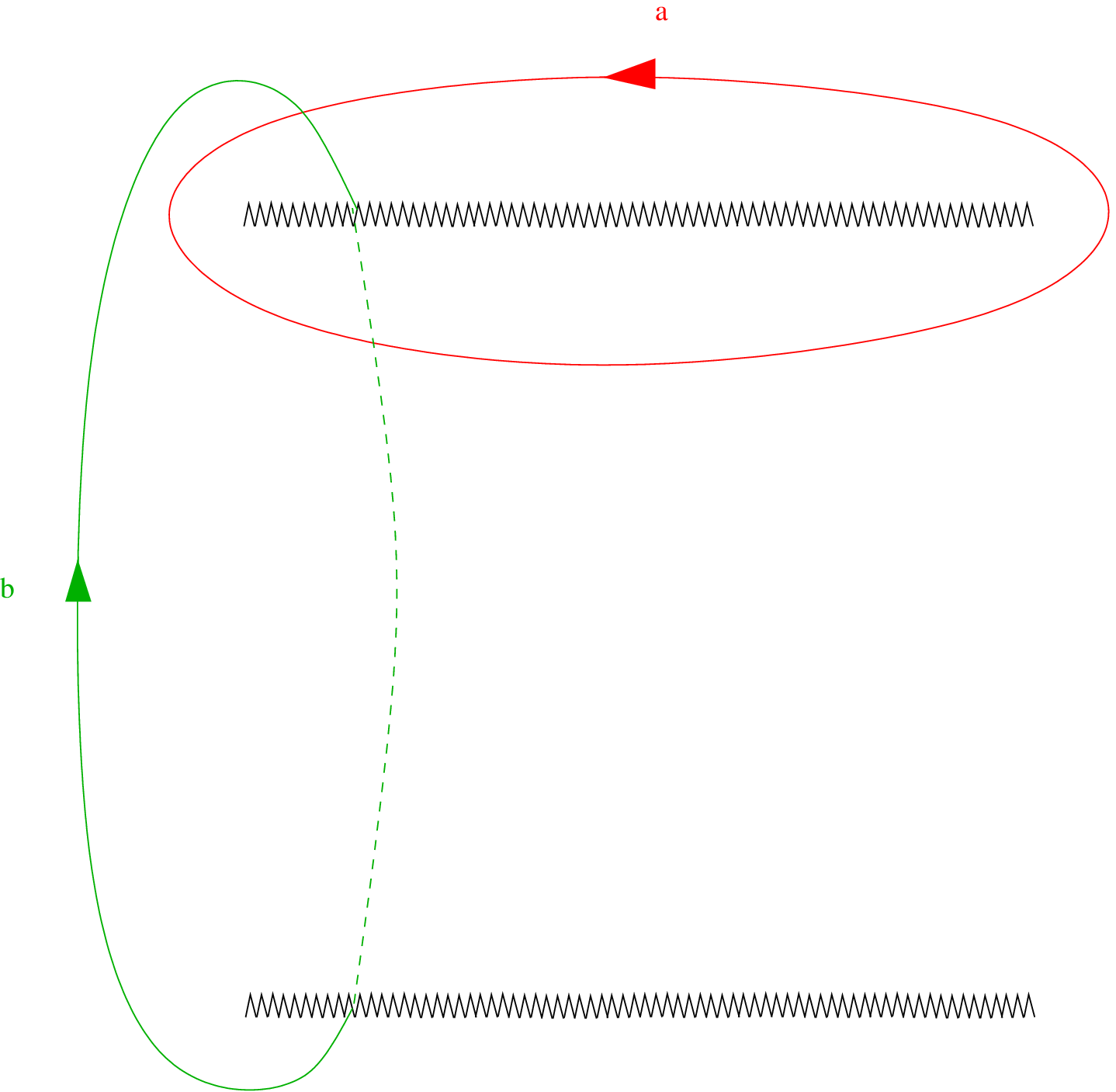}\\
$(a)$ & & $(b)$ & & $(c)$
\end{tabular}
\caption{$(a)$ The elliptic curve $\Sigma$ with its 4 branch
points and its canonical $a$- and $b$-cycles of $H_1(\Sigma)$.
$(b)$ The same elliptic curve after applying the automorphism
which has the following action $a \mapsto b, b \mapsto -a$ on
$H_1(\Sigma)$. Note that the branch points are unchanged since the
curve is mapped to itself. $(c)$ Since the branch cuts are purely
a matter of choice, it is convenient to redefine them so as to
make the new $a$- and $b$-cycles take their standard form on
$\Sigma$.} \label{a b cycles}
\end{figure}

\begin{figure}[h]
\centering \psfrag{a}{\textcolor{red}{\tiny $a$}}
\psfrag{b}{\textcolor{green}{\tiny $b$}}
\psfrag{c}{\textcolor{blue}{\tiny $c$}}
\includegraphics[width=27mm]{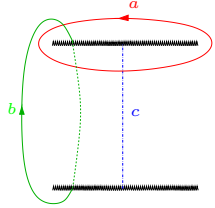}
\caption{Artificial `condensate cut' $c$ for $dp$.}
\label{condensate cut}
\end{figure}

Recall that in the finite-gap integration method we have an
elliptic curve $\Sigma$ equipped with a normalised meromorphic
$1$-form $dp$ with the following periods
\begin{equation*}
\int_a dp = 0, \quad \int_b dp = 2 \pi n, \quad n \in \mathbb{Z}.
\end{equation*}
The vanishing of the $a$-periods of $dp$ guarantees that the
Abelian integral $p(x) = \int^{x^+} dp$ is a well define and
single valued function on the top sheet. If the $a$-periods of
$dp$ were non-vanishing, a condensate cut would be needed on
$\Sigma$ in order to define a single valued domain of the
differential $dp$ on the top sheet of $\Sigma$. In this sense
condensate cuts are due to non-zero $a$-periods of $dp$ (the
terminology comes from the gauge theory side; see for instance
\cite{KMMZ}). It is important to note however that condensate cuts
are not a property of the elliptic curve $\Sigma$ but rather of
the Abelian differential $dp$ living on $\Sigma$. In order to
obtain the condensate cut picture of giant magnons of
\cite{Minahan:2006bd}, one could at this stage apply the modular
transformation $\tau \mapsto - \frac{1}{\tau}$ which has the
effect of swapping around the $a$- and $b$-cycles as $a \mapsto b,
b \mapsto -a$ (see Figure \ref{a b cycles}) so that the
differential $dp$ now satisfies
\begin{equation*}
\int_b dp = 0, \quad \int_a dp = 2 \pi n, \quad n \in \mathbb{Z}.
\end{equation*}
Notice that the Abelian integral $p(x) = \int^{x^+} dp$ does not
define a single valued function on the top sheet anymore. One must
introduce a `condensate cut' $c$ on the top sheet to define a
single branch of this multivalued function. This cut should
intersect the $a$-cycle once, which can be achieved by connecting
the branch cuts of $\Sigma$ (see Figure \ref{condensate cut}).

Shrinking the $b$-period in the original picture of Figure \ref{a
b cycles} $(a)$ now corresponds to shrinking the $a$-period in
Figure \ref{condensate cut} to a point. In particular, the $k
\rightarrow 1$ limit of the curve which takes $x_1 \rightarrow
x_2$ (and $\bar{x}_1 \rightarrow \bar{x}_2$) leaves behind a
condensate cut on $\Sigma$, connecting the two singular points
$x_1 = x_2, \bar{x}_1 = \bar{x}_2$. This picture of course is no
different to the one we used in section \ref{section: HM limit},
where the non-vanishing $b$-cycle around $x_1 = x_2$ could equally
be interpreted as a `condensate' connecting $x_1 = x_2$ to
$\bar{x}_1 = \bar{x}_2$. A multivalued function $f$ on the top
sheet can be described in two equivalent ways: either using
condensate cuts or by stating that $\int_c df \neq 0$ for some
closed curve $c$ lying entirely on the top sheet. In the case of the
quasi-momentum $p(x)$, the conditions that $\oint_{x_1} dp \neq 0$ and
$\oint_{\bar{x}_1} dp \neq 0$ can be accounted for by requiring $p(x)$
to have simple poles at $x_1$ and $\bar{x}_1$, as was done in
\cite{Minahan:2006bd}.

\section*{Acknowledgements}
I would like to thank Keisuke Okamura and Ryo Suzuki for interesting
discussions. This work is supported by EPSRC.

\appendix

\section{Elliptic setup} \label{section: Elliptic setup}

Throughout these appendices we will be relying heavily on the book by
Byrd and Friedman \cite{B&M} for computations of elliptic integrals,
and so we adopt the convention that any equation reference of the form
``xxx.yy'' refers to an equation in this book. Part of the
calculation parallels that in \cite{Filament}.

\begin{figure}
\centering \psfrag{0}{\footnotesize{$0$}}
\psfrag{1}{\footnotesize{$1$}}
\psfrag{pp}{\footnotesize{$(k')^{-2}$}}
\psfrag{a}{\footnotesize{$a$}} \psfrag{b}{\footnotesize{$b$}}
\psfrag{w}{\tiny{$w$}}
\includegraphics[width=50mm]{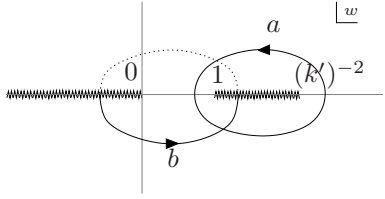}
\caption{$a$- and $b$-periods in $w$-plane.} \label{a b periods 2}
\end{figure}

In order to simplify things, we first seek the
M\"obius transformation that sends three of the branch points to
the real axis, for instance $x_2 \rightarrow 0, x_1 \rightarrow 1,
\bar{x}_2 \rightarrow \infty$, which is easily found to be
\begin{equation} \label{Mobius}
w(x) = \frac{x - x_2}{h(x - \bar{x}_2)}, \qquad h \equiv
\frac{x_1 - x_2}{x_1 - \bar{x}_2}.
\end{equation}
In the $w$-plane the branch point $\bar{x}_1$ gets mapped to the real
line
\begin{equation*}
w(\bar{x}_1) = \frac{1}{|h|^2} = (k')^{-2}, \qquad k' \equiv |h| \in
\mathbb{R},
\end{equation*}
as it should be since the branch points $x_1, \bar{x}_1, x_2,
\bar{x}_2$ are concentric in the $x$-plane. Note that $k' < 1$
follows from our choice of putting both roots $x_1, x_2$ in the
upper-half of the $x$-plane; see Figure \ref{a b periods 2}.

\begin{figure}
\centering \psfrag{1}{\footnotesize{$1$}}
\psfrag{pp}{\footnotesize{$(k')^{-2}$}}
\psfrag{a1}{\textcolor{red}{\footnotesize{$a_1$}}}
\psfrag{b1}{\textcolor{green}{\footnotesize{$b_1$}}}
\psfrag{x}{\tiny{$x$}}
\psfrag{x1}{\tiny{$x_1$}}
\psfrag{x2}{\tiny{$x_2$}}
\psfrag{xb1}{\tiny{$\bar{x}_1$}}
\psfrag{xb2}{\tiny{$\bar{x}_2$}}
\psfrag{w}{\tiny{$w$}}
\includegraphics[width=120mm]{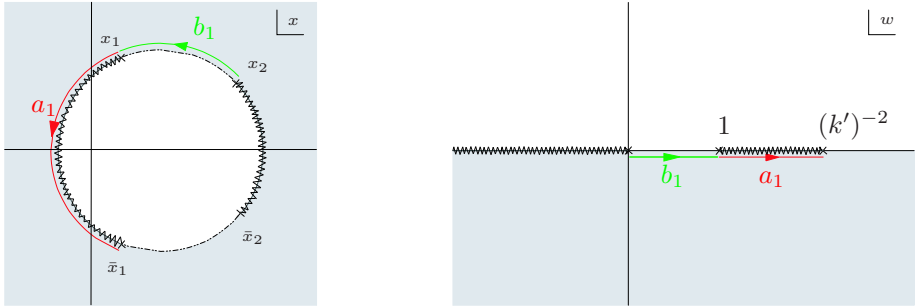}
\caption{Choice of branch for $y$.} \label{y branch}
\end{figure}

The inverse of \eqref{Mobius} is given by
\begin{equation} \label{Mobius 2}
x(w) = \frac{x_2 - \bar{x}_2 h  w}{1 - h w},
\end{equation}
and so given any $x_0 \neq \bar{x}_2$ such that $w_0 = w(x_0) \neq
\infty$ one can show that
\begin{subequations} \label{x -> w}
\begin{equation} \label{x -> w 1}
x - x_0 = \frac{h (x_2 - \bar{x}_2) (w - w_0)}{(1 - h w) (1 - h
w_0)},
\end{equation}
and in the limit $w_0 \rightarrow \infty$ we obtain also
\begin{equation} \label{x -> w 2}
x - \bar{x}_2 = \frac{x_2 - \bar{x}_2}{1 - h w}.
\end{equation}
\end{subequations}
Using \eqref{x -> w} and after a little algebra we can reexpress
the curve \eqref{curve} in the $w$-coordinate
\begin{equation} \label{curve2}
y^2 = -h^2 |x_1 - \bar{x}_2|^2 \left( \frac{x_2 - \bar{x}_2}{(1 - h
w)^2} \right)^2 w (w - 1)(1 - (k')^2 w).
\end{equation}
Also, equation \eqref{Mobius 2} can be used to derive
\begin{equation} \label{d Mobius}
dx = \frac{h (x_2 - \bar{x}_2)}{(1 - h w)^2} dw.
\end{equation}
Before proceeding, we must first specify a branch for the function $y$
that we will be working with. We do this by requiring that $\text{Re}
\; y > 0$ along the path $a_1$ in Figure \ref{y branch}, which lies
just below the cut from $1$ to $(k')^{-2}$ in the $w$-plane.
As we move along the path $a_1$ we have $\text{Im} (dx) < 0$ and
$\text{Re} (dw) > 0$ so that \eqref{d Mobius} implies
\begin{equation*}
\text{Re} \; \left( \frac{h}{(1 - h w)^2} \right) < 0,
\end{equation*}
and since $w (w - 1)(1 - (k')^2 w) > 0$ for $w \in [1, (k')^{-2} ]$ we find
\begin{equation} \label{y}
y = - \left| (x_1 - \bar{x}_2)(x_2 - \bar{x}_2) \right| \frac{h}{(1 - h
w)^2} \sqrt{w (w - 1)(1 - (k')^2 w)},
\end{equation}
where $z \rightarrow \sqrt{z}$ denotes the principal branch of the
square root (i.e. the branch $\text{arg} z \in (-\pi, \pi]$ on
which $\text{Re} \sqrt{z} > 0$). We can now obtain a simple
expression for the unique (up to a multiplicative constant)
holomorphic differential on the elliptic curve, namely
\begin{equation} \label{nu}
\nu \equiv \frac{dx}{y} = \frac{-i}{|x_1 - \bar{x}_2|}
\frac{dw}{\sqrt{w (w - 1)(1 - (k')^2 w)}}.
\end{equation}

Expressed in the $w$-coordinate it is simple to obtain the $a$-
and $b$-periods of $\nu$ in terms of elliptic functions. For
instance, since the value of \eqref{y} has opposite sign on either
side of the cut from $1$ to $(k')^{-2}$, the $a$-period is just twice
the integral along the curve $a_1$ in Figure \ref{y branch}, which
using 236.00 can be expressed in terms of standard elliptic functions
\begin{equation*}
\int_a \nu = \frac{- 2i}{|x_1 - \bar{x}_2|} \int_1^{(k')^{-2}}
\frac{dw}{\sqrt{w (w - 1)(1 - (k')^2 w)}} = \frac{- 4 i K}{|x_1 -
\bar{x}_2|}, \qquad K \equiv K(k),
\end{equation*}
where $k \equiv \sqrt{1 - (k')^2}$ is the complementary modulus to
$k'$. It is thus natural to define the unique normalised
holomorphic differential as follows
\begin{equation} \label{omega def}
\omega \equiv - \frac{|x_1 - \bar{x}_2|}{4 i K} \nu.
\end{equation}
Similarly the $b$-period is twice the integral along the curve
$b_1$ lying just below the real axis as shown in Figure \ref{y
branch}. However, one must be careful in dealing with the branch
of the square root, so that for instance one has
\begin{equation*}
\sqrt{w (w - 1)(1 - (k')^2 w)} = - i \sqrt{w (1 - w)(1 - (k')^2 w)}
\quad  \text{for} \; w \in b_1.
\end{equation*}
Equation 234.00 then leads to
\begin{equation*}
\int_b \nu = \frac{-2i}{|x_1 - \bar{x}_2|} \int_0^1 \frac{dw}{-i \sqrt{w (1
- w)(1 - (k')^2 w)}} = \frac{4 K'}{|x_1 - \bar{x}_2|}, \qquad K' \equiv
K(k').
\end{equation*}
The period matrix in the elliptic case is simply a number given
here by
\begin{equation*}
\Pi = \int_b \omega = - \frac{|x_1 - \bar{x}_2|}{4 i K} \int_b \nu =
\frac{i K'}{K} \equiv \tau.
\end{equation*}

\section{$\Theta$-functions} \label{section: Theta functions}

To compute $\int_{\infty^{\pm}}^{0^{\pm}} \omega$ which appear inside
the $\theta$-functions appearing in \eqref{reconstruction formula for
psi^+} we will evaluate the integrals $\int_{\infty^-}^{\infty^+}
\omega$ and $\int_{0^-}^{0^+} \omega$ since
\begin{equation} \label{int rel}
\int_{\infty^{\pm}}^{0^+} \omega = \frac{1}{2} \left(
\int_{0^-}^{0^+} \omega \mp \int_{\infty^-}^{\infty^+} \omega
\right).
\end{equation}
Such a decomposition is useful because an integral of the form
$\int_{x^-}^{x^+} \omega$ can be related to the $b$-periods of a
normalised Abelian differential of the third kind $\omega_x$ with
only poles at $x^{\pm}$ and defined uniquely by its residues there
\begin{equation} \label{3rd kind diff}
\int_a \omega_x = 0, \qquad \text{res}_{x^{\pm}} \omega_x = \pm
\frac{1}{2 \pi i}.
\end{equation}
More precisely, what one can show using the Riemann bilinear
identities is that
\begin{equation*}
\int_{x^-}^{x^+} \omega = \int_b \omega_x.
\end{equation*}
To compute the $b$-periods of $\omega_0$ and $\omega_{\infty}$ we
need explicit expressions for these differentials. However, since
they are both uniquely defined by the conditions \eqref{3rd kind
diff}, if we can explicitly construct differentials satisfying
\eqref{3rd kind diff} we are done. But $-\frac{x}{2 \pi i} \nu$ has
the right residues\footnote{This is because $-\frac{x}{2 \pi i} \nu =
-\frac{x}{2 \pi i} \frac{dx}{y}$ and $y \sim_{x \rightarrow \infty} x^2$ (since
$\text{Re} \, y(x) > 0$ for large $x \in \mathbb{R}$ by our choice
of branch for $y$) so that $-\frac{x}{2 \pi i} \nu = -\frac{1}{2 \pi
i} \frac{dx}{x} = \frac{1}{2 \pi i} \frac{d\lambda}{\lambda}$ where
$\lambda = \frac{1}{x}$ is a local parameter near $x = \infty$.} at $x
= \infty$ and $\frac{y(0)}{2 \pi i x} \nu$ the right residues at $x =
0$ so all we need to do is normalise these differentials
(i.e. subtract multiples of the holomorphic differential $\nu$ to make
their $a$-periods vanish), so that
\begin{equation} \label{omega inf and zero}
\omega_{\infty} = - \frac{1}{2 \pi i} \left( x - L_{\infty} \right)
\nu, \qquad \omega_0 = \frac{y(0)}{2 \pi i} \left( \frac{1}{x} -
L_0 \right) \nu,
\end{equation}
where we have defined
\begin{equation} \label{L def}
L_{\infty} \equiv \frac{\int_a x \nu}{\int_a \nu}, \qquad L_0
\equiv \frac{\int_a \frac{1}{x} \nu}{\int_a \nu}.
\end{equation}
Using equations \eqref{x -> w 2} and \eqref{nu} we can write
\begin{equation*}
\begin{split}
\int_b x \nu &= \frac{-2i}{|x_1 - \bar{x}_2|} \int_0^1 \left(
\bar{x}_2 + \frac{x_2 - \bar{x}_2}{1 - h w} \right) \frac{dw}{-i
\sqrt{w (1 - w)(1 - (k')^2 w)}},\\
\int_b \frac{1}{x} \nu &= \frac{-2i}{|x_1 - \bar{x}_2|} \int_0^1
\left( \frac{1}{\bar{x}_2} + \frac{1}{\bar{x}_2} \frac{x_2 -
\bar{x}_2}{\bar{x}_2 h w - x_2} \right) \frac{dw}{-i \sqrt{w (1 -
w)(1 - (k')^2 w)}}.
\end{split}
\end{equation*}
The second summands in both of these integrals can be reduced to
canonical elliptic functions using 233.02, whereas the first
summands are proportional to $\int_b \nu$ which has already been
computed. So we end up with
\begin{subequations} \label{int omega}
\begin{equation} \label{int omega 1}
\int_{\infty^-}^{\infty^+} \omega = \int_b \omega_{\infty} =
\frac{- 1}{2 \pi i}\frac{4}{|x_1 - \bar{x}_2|} \left[ (\bar{x}_2 -
L_{\infty}) K' + (x_2 - \bar{x}_2) \Pi(h,k') \right],
\end{equation}
\begin{equation} \label{int omega 2}
\int_{0^-}^{0^+} \omega = \int_b \omega_0 = \frac{y(0)}{2 \pi
i}\frac{4}{|x_1 - \bar{x}_2|} \left[ \left(\frac{1}{\bar{x}_2} -
L_0\right) K' + \left(\frac{1}{x_2} - \frac{1}{\bar{x}_2}\right)
\Pi\left( k ,k'\right) \right],
\end{equation}
\end{subequations}
where $h_2 \equiv h \frac{\bar{x}_2}{x_2} = w(0)^{-1}$, $w(x)$ being the
M\"obius transformation \eqref{Mobius} to the $w$-plane. Although these
integrals look complicated, it turns out that their real parts are
very simple. They can be computed using the addition formula 117.02
for elliptic functions of the third kind $\Pi\left(h,k'\right)$, after
noting the relations $\bar{h} = (k')^2/h$ and $\bar{h}_2 = (k')^2 /
h_2$ between their arguments.

If we assume that the two cuts are on either side of the origin $x
= 0$ then it follows from our choice of branch for $y$ that $y(0)
= - |x_1| |x_2| < 0$: indeed, all along the real axis $y \in
\mathbb{R}$, but $y(x) > 0$ when $x$ lies outside the region in
between the two cuts, and $y(x) < 0$ when $x$ lies between the two
cuts. After some algebra we find
\begin{equation*}
\text{Re} \left( \int_{\infty^-}^{\infty^+} \omega \right) =
\frac{1}{2} \int_{\infty^-}^{\infty^+} ( \omega + \bar{\omega} ) =
- \frac{1}{2}, \qquad \text{Re} \left( \int_{0^-}^{0^+} \omega
\right) = \frac{1}{2} \int_{0^-}^{0^+} ( \omega + \bar{\omega} ) =
\frac{1}{2}.
\end{equation*}
Equation \eqref{rho def} in the text now follows immediately from this
and equation \eqref{int rel}.

We now verify that the variables $\rho_{\pm} \in \mathbb{R}$ in
\eqref{rho def} together represent an extra degree of freedom of the
solution that cannot be obtained simply from the modulus $k'$ of the
curve. For this we will find simpler expressions for \eqref{int
  omega}. The quantities $L_{\infty}, L_0$ defined in \eqref{L def}
can be computed explicitly using 236.00, 236.02 and the fact that
$L_{\infty}, L_0 \in \mathbb{R}$ (or equivalently using 117.03); after
some algebra one finds
\begin{subequations} \label{L}
\begin{equation} \label{L inf}
L_{\infty} = x_2 + (x_2 - \bar{x}_2) \frac{h}{1 - h}
\frac{\Pi(\beta^2,k)}{K}, \qquad \beta^2 \equiv \frac{k^2}{1 - h}.
\end{equation}
\begin{equation} \label{L zero}
L_0 = \frac{1}{x_2} + \left(\frac{1}{x_2} -
\frac{1}{\bar{x}_2}\right) \frac{h_2}{1 - h_2}
\frac{\Pi\left((\beta')^2,k\right)}{K}, \qquad (\beta')^2 \equiv
\frac{k^2}{1 - h_2}.
\end{equation}
\end{subequations}
Now using the addition formula 117.05 we can relate $\Pi(h,k')$
which appears in \eqref{int omega 1} to $\Pi(\beta^2,k)$ which
appears in \eqref{L inf} on the one hand and $\Pi(h_2,k')$ which
appears in \eqref{int omega 2} to $\Pi((\beta')^2,k)$ which
appears in \eqref{L zero} on the other
\begin{subequations} \label{useful rel 1}
\begin{equation} \label{useful rel 1-1}
(h - 1) K \Pi(h,k') + h K' \Pi\left( \beta^2, k \right) + (1 - h)
K K' = \frac{\pi}{2} \sqrt{\frac{1-h}{\bar{h} - 1}} F(\phi, k'),
\end{equation}
\begin{equation} \label{useful rel 1-2}
(h_2 - 1) K \Pi(h_2,k') + h_2 K' \Pi\left( (\beta')^2, k \right) +
(1 - h_2) K K' = \frac{\pi}{2} \sqrt{\frac{1-h_2}{\bar{h}_2 - 1}}
F(\phi', k'),
\end{equation}
\end{subequations}
where
\begin{equation} \label{phi & phi' def}
\phi \equiv \text{sin}^{-1}\left( \frac{\sqrt{h}}{k'} \right),
\qquad \phi' \equiv \text{sin}^{-1}\left( \frac{\sqrt{h_2}}{k'}
\right).
\end{equation}
Note that each square root in these expressions is taken to be the
principal branch of the square root as previously defined.
Equations \eqref{useful rel 1} lead to great simplifications in
\eqref{int omega}, and after the dust settles we end up with
\begin{subequations} \label{int omega final}
\begin{equation} \label{int omega final 1}
\int_{\infty^-}^{\infty^+} \omega = - i \frac{x_2 -
\bar{x}_2}{|x_1 - \bar{x}_2|} \frac{1}{1 - h} \sqrt{\frac{1 -
h}{\bar{h} - 1}} \frac{F(\phi, k')}{K},
\end{equation}
\begin{equation} \label{int omega final 2}
\int_{0^-}^{0^+} \omega = - i \frac{y(0)}{|x_2|^2} \frac{x_2 -
\bar{x}_2}{|x_1 - \bar{x}_2|} \frac{1}{1 - h_2} \sqrt{\frac{1 -
h_2}{\bar{h}_2 - 1}} \frac{F(\phi', k')}{K}.
\end{equation}
\end{subequations}
To simplify this expression further one must be careful in dealing
with the square roots. We must deal separately with two cases:
$(1)$ when $x = 0$ lies between the two cuts on the one hand, and
$(2)$ when $x = 0$ lies outside the two cuts on the other (we deal
with this last case as well for completeness).

\begin{figure}
\centering \psfrag{x}{\tiny{$x$}} \psfrag{x1}{\tiny{$x_1$}}
\psfrag{x2}{\tiny{$x_2$}} \psfrag{xb1}{\tiny{$\bar{x}_1$}}
\psfrag{xb2}{\tiny{$\bar{x}_2$}} \psfrag{0}{\tiny{$0$}}
\includegraphics[width=40mm]{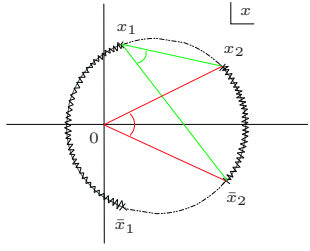}
\caption{$y(0) < 0 \Leftrightarrow \text{Arg} \, h_2 \in (-\pi,
0)$.} \label{double angle}
\end{figure}

In case $(1)$ we have $y(0) < 0 $ and it is not too hard to show
that $\text{Arg} \, h_2 \in (-\pi, 0)$ (see Figure \ref{double
angle}) so that $\text{Arg} \left((1 - h_2)/(\bar{h}_2 - 1)
\right) \in (-\pi, 0)$ which implies
\begin{equation*}
\underline{\text{case} \; (1):} \quad \frac{1}{1 - h_2}
\sqrt{\frac{1 - h_2}{\bar{h}_2 - 1}} = \frac{- i}{|1 - h_2|}.
\end{equation*}
In case $(2)$ we have $y(0) > 0 $ and now one can show that
$\text{Arg} \, h_2 \in (0, \pi)$ (see Figure \ref{double angle})
so that this time we have $\text{Arg} \left((1 - h_2)/(\bar{h}_2 -
1) \right) \in (0, \pi)$ which implies
\begin{equation*}
\underline{\text{case} \; (2):} \quad \frac{1}{1 - h_2}
\sqrt{\frac{1 - h_2}{\bar{h}_2 - 1}} = \frac{i}{|1 - h_2|}.
\end{equation*}
Thus, in both cases we can write
\begin{equation*}
\frac{1}{1 - h} \sqrt{\frac{1 - h}{\bar{h} - 1}} = \frac{i}{|1 -
h|}, \qquad \frac{y(0)}{1 - h_2} \sqrt{\frac{1 - h_2}{\bar{h}_2 -
1}} = \frac{i |x_1||x_2|}{|1 - h_2|}.
\end{equation*}
Plugging these expressions back into the equations \eqref{int
omega final} we finally obtain
\begin{subequations} \label{int omega final 2}
\begin{equation} \label{int omega final 2 1}
\int_{\infty^-}^{\infty^+} \omega = i \frac{F(\phi, k')}{K},
\end{equation}
\begin{equation} \label{int omega final 2 2}
\int_{0^-}^{0^+} \omega = i \frac{F(\phi', k')}{K}.
\end{equation}
\end{subequations}
We are now in a position to obtain explicit expressions for
$\rho_{\pm}$ by making use of formula 116.01 to obtain the sum and the
difference of \eqref{int omega final 2 1} and \eqref{int omega final 2
2} given in \eqref{int rel}. The result is given in \eqref{int rel
final} and \eqref{phi final 2}, where the fact that $\sqrt{\bar{h} -
1} = - i \sqrt{1 - \bar{h}}$ and $\sqrt{\bar{h}_2 - 1} = i \sqrt{1 -
\bar{h}_2}$ has been used to obtain \eqref{phi final 2}.

\section{Exponentials} \label{section: Exponentials}

For the evaluation of the integrals $\int_{\infty^{\pm}}^{0^+} d
\mathcal{Q}$ this we will start by evaluating the integrals
$\int_{\infty^-}^{\infty^+} d \mathcal{Q}$ and $\int_{0^-}^{0^+} d
\mathcal{Q}$ since
\begin{equation*}
\int_{\infty^{\pm}}^{0^+} d \mathcal{Q} = \frac{1}{2} \left(
\int_{0^-}^{0^+} d \mathcal{Q} \mp \int_{\infty^-}^{\infty^+} d
\mathcal{Q} \right).
\end{equation*}
As in \eqref{int rel} we can relate the integral $\int_{x^-}^{x^+} d
\mathcal{Q}$ for general $x$ to the third kind Abelian differential
$\omega_x$ defined by \eqref{3rd kind diff}. To do this we will again
make use of the Riemann bilinear identities but this time for the
differentials $dp$ (or $dq$) and $\omega_x$, thus obtaining
\begin{equation*}
\int_{x^-}^{x^+} dp = - \sum \text{res} \; p (2 \pi i \omega_x),\qquad
\int_{x^-}^{x^+} dq = - \sum \text{res} \; q (2 \pi i \omega_x).
\end{equation*}
Making use of equation \eqref{omega inf and zero} for
$\omega_{\infty}$ and $\omega_0$ as well as \eqref{p asymp} and
\eqref{q asymp} we can evaluate the residues in the last equations
explicitly so that the integrals of $d\mathcal{Q}$ reduce to
\begin{subequations} \label{int dQ}
\begin{equation} \label{int dQ 1}
\int_{\infty^+}^{0^+} d\mathcal{Q} = - \frac{1 + y(0)}{|x_1 -
\bar{x}_2|} \tilde{T} + \frac{L_{\infty} + y(0) L_0}{|x_1 -
\bar{x}_2|} \left( \tilde{X} - \tilde{X}_0 \right),
\end{equation}
\begin{equation} \label{int dQ 2}
\int_{\infty^-}^{0^+} d\mathcal{Q} = \frac{1 - y(0)}{|x_1 -
\bar{x}_2|} \tilde{T} - \frac{L_{\infty} - y(0) L_0}{|x_1 -
\bar{x}_2|} \left( \tilde{X} - \tilde{X}_0 \right).
\end{equation}
\end{subequations}
Now using equations 117.05 and 110.10 (and with the help of
\eqref{useful rel 1}) one can show that the following hold
\begin{subequations} \label{Pi to Zeta 2}
\begin{equation} \label{Pi to Zeta 2 1}
\frac{\Pi(\beta^2,k)}{K} = \frac{1}{h} \sqrt{\frac{1 - h}{\bar{h}
- 1}} \left[ Z_0(\phi,k') + \frac{\pi}{2} \frac{F(\phi,k')}{K K'}
\right],
\end{equation}
\begin{equation} \label{Pi to Zeta 2 2}
\frac{\Pi \left( (\beta')^2,k \right)}{K} = \frac{1}{h_2}
\sqrt{\frac{1 - h_2}{\bar{h}_2 - 1}} \left[ Z_0(\phi',k') +
\frac{\pi}{2} \frac{F(\phi',k')}{K K'} \right],
\end{equation}
\end{subequations}
where $\phi, \phi'$ were defined in \eqref{phi & phi' def},
$\beta^2, (\beta')^2$ are defined in \eqref{useful rel 1} and
$Z_0(\theta,k')$ is the Jacobi zeta-function defined as
\begin{equation*}
Z_0(\theta,k') = E(\theta,k') - \frac{E'}{K'} F(\theta,k').
\end{equation*}
We can now use \eqref{Pi to Zeta 2} to obtain
\begin{subequations} \label{L2}
\begin{equation} \label{L2 inf}
L_{\infty} = x_2 - |x_1 - \bar{x}_2| \left[ Z_0(\phi,k') +
\frac{\pi}{2} \frac{F(\phi,k')}{K K'} \right],
\end{equation}
\begin{equation} \label{L2 zero}
y(0) L_0 = \frac{y(0)}{x_2} + |x_1 - \bar{x}_2| \left[
Z_0(\phi',k') + \frac{\pi}{2} \frac{F(\phi',k')}{K K'} \right].
\end{equation}
\end{subequations}
Using the addition formula 142.01 for Jacobi zeta-functions we
find
\begin{equation*}
\frac{L_{\infty} \mp y(0) L_0}{|x_1 - \bar{x}_2|} = \frac{x_2 \mp
\frac{y(0)}{x_2}}{|x_1 - \bar{x}_2|} - \sqrt{h} \sqrt{h_2} \sin
\varphi_{\pm} \mp \left[ Z_0(\varphi_{\pm},k') + \frac{\pi}{2}
\frac{F(\varphi_{\pm},k')}{K K'} \right].
\end{equation*}
Now using 143.02 for Jacobi zeta-functions with imaginary
arguments (and also 126.01 for Jacobian elliptic functions with
imaginary arguments) we can write the expression in square
brackets in the last expression as
\begin{equation*}
\left[ Z_0(u_{\pm},k') + \frac{\pi}{2} \frac{u_{\pm}}{K K'}
\right] = i Z_0(i u_{\pm}, k) + \tan \varphi_{\pm} \sqrt{1 -
(k')^2 \sin^2 \varphi_{\pm}},
\end{equation*}
where $u_{\pm} = F(\varphi_{\pm},k') =
\text{am}^{-1}(\varphi_{\pm},k')$, and $Z_0(u_{\pm},k') =
Z_0(\varphi_{\pm},k')$ is just a standard change in notation of
the argument. Grouping everything together we obtain,
\begin{equation} \label{exponent + sign X}
\mp \frac{L_{\infty} \mp y(0) L_0}{|x_1 - \bar{x}_2|} = i
Z_0\left(i u_{\pm}, k\right) \mp \left[ \frac{x_2 \mp
\frac{y(0)}{x_2}}{|x_1 - \bar{x}_2|} - \sqrt{h} \sqrt{h_2} \sin
\varphi_{\pm} \mp \tan \varphi_{\pm} \sqrt{1 - (k')^2 \sin^2
\varphi_{\pm}} \right].
\end{equation}
Everything in the square bracket of this last expression can be
expressed in terms of the branch points $\{ x_1, \bar{x}_1, x_2,
\bar{x}_2 \}$ by making use of \eqref{phi final 2},
\begin{equation*}
\begin{split}
\tan \varphi_{\pm} &= \mp \frac{|x_1 - \bar{x}_2|}{\sqrt{x_1 x_2} \pm
\sqrt{\bar{x}_1 \bar{x}_2}},\\
\sin \varphi_{\pm} &= \frac{|x_1 - \bar{x}_2|}{|x_2| \mp |x_1|},\\
\sqrt{ 1 - (k')^2 \sin^2 \varphi_{\pm}} &= \pm
\frac{\sqrt{\bar{x}_1 x_2} \pm \sqrt{x_1 \bar{x}_2}}{|x_2| \pm
|x_1|}.
\end{split}
\end{equation*}
Note that to obtain the right sign in the last of these expressions
one needs to use the fact that $\sqrt{x_1 \bar{x}_2} + \sqrt{\bar{x}_1
x_2}$ is real and positive. We also assume that $|x_1| \geq
|x_2|$ and that $x = 0$ lies between the two cuts. After simplifying
the expression in the square bracket of \eqref{exponent + sign X} one
finds (including the overall ``$\mp$'' sign sitting outside the square
bracket)
\begin{equation*}
- \frac{\left( |x_2| \pm |x_1| \right) \left( \sqrt{\bar{x}_1 x_2}
\pm \sqrt{x_1 \bar{x}_2} \right)}{|x_1 - \bar{x}_2| \left(
\sqrt{x_1 x_2} \pm \sqrt{\bar{x}_1 \bar{x}_2} \right)} =
\frac{\tan \varphi_{\pm} \sqrt{ 1 - (k')^2 \sin^2
\varphi_{\pm}}}{\sin^2 \varphi_{\pm}} = i \, \text{cs}(i
u_{\pm},k) \text{dn}(i u_{\pm},k),
\end{equation*}
where use was made of 125.02 for Jacobi elliptic functions with
imaginary arguments. Now one of the equations among 141.01 implies the
following
\begin{equation*}
Z_0(i u_{\pm},k) + \text{cs}(i u_{\pm},k) \text{dn}(i u_{\pm},k) =
Z_0(i u_{\pm} + i K',k) + \frac{i \pi}{2 K} \equiv Z_1(i
u_{\pm},k),
\end{equation*}
where the last equality is the definition of the Jacobi
zeta-function $Z_1$. So finally,
\begin{equation} \label{exponent in terms of Z_1}
\mp \frac{L_{\infty} \mp y(0) L_0}{|x_1 - \bar{x}_2|} = i Z_1(i
u_{\pm}, k).
\end{equation}
Using $u_+ = F(\varphi_+,k') = \tilde{\rho}_- + K'$, equation
\eqref{exponent in terms of Z_1} gives in the ``$+$'' sign case
\begin{equation*}
- \frac{L_{\infty} - y(0) L_0}{|x_1 - \bar{x}_2|} = i Z_1(i u_+,
k) = i Z_1(i \tilde{\rho}_- + K', k).
\end{equation*}
Now plugging this back into \eqref{int dQ 2} yields
\begin{equation*}
-i \int_{\infty^-}^{0^+} d\mathcal{Q} = - i \frac{1 - y(0)}{|x_1 -
\bar{x}_2|} \tilde{T} + Z_1(i \tilde{\rho}_- + K', k) \left(
\tilde{X} - \tilde{X}_0 \right),
\end{equation*}
and since $\frac{i}{2} \int_b d\mathcal{Q} = \frac{i \pi}{2 K}
(\tilde{X} - \tilde{X}_0)$ it follows that the exponent in
\eqref{reconstruction formula for Z3_2} can be written
\begin{equation} \label{int dQ infty-}
-i \int_{\infty^-}^{0^+} d\mathcal{Q} + \frac{i}{2} \int_b
d\mathcal{Q} = Z_0(i \tilde{\rho}_-,k) \left( \tilde{X} -
\tilde{X}_0 \right) - i \frac{1 - y(0)}{|x_1 - \bar{x}_2|}
\tilde{T}.
\end{equation}

In the ``$-$'' sign case, using $i u_- = i F(\varphi_-,k') = K + i
\tilde{\rho}_+$, equation \eqref{exponent in terms of Z_1} yields
\begin{equation*}
\frac{L_{\infty} + y(0) L_0}{|x_1 - \bar{x}_2|} = i Z_1(K + i
\tilde{\rho}_+, k) = i Z_2(i \tilde{\rho}_+, k),
\end{equation*}
where the Jacobi zeta-functions $Z_2$ is defined as
\begin{equation*}
Z_2(u,k) = Z_1(u + K,k).
\end{equation*}
Equation \eqref{int dQ 1} then yields
\begin{equation} \label{int dQ infty+}
-i \int_{\infty^+}^{0^+} d\mathcal{Q} = Z_2(i \tilde{\rho}_+,k)
\left( \tilde{X} - \tilde{X}_0 \right) + i \frac{1 + y(0)}{|x_1 -
\bar{x}_2|} \tilde{T}.
\end{equation}

Defining the following variables
\begin{equation*}
v_{\pm} = \frac{y(0) \pm 1}{|x_1 - \bar{x}_2|},
\end{equation*}
appearing in \eqref{int dQ infty-} and \eqref{int dQ infty+}, it can
be shown (using 122.03, 122.05 and 125.02) that they satisfy
\begin{equation*}
v_-^2 - v_+^2 = \text{dn}^2 (i \tilde{\rho}_-,k) + (k')^2
\text{sc}^2 (i \tilde{\rho}_+,k).
\end{equation*}

\section{Normalisation} \label{section: Normalisation}

Let us now turn to the evaluation of the constants $h_{\pm}(0^+)$
and $\chi(\infty^-)^{\frac{1}{2}}$ in the elliptic case at hand.
Recall that $h_{\pm}$ are meromorphic functions on $\Sigma$
uniquely defined (up to normalisation) by having poles at the
divisor $\hat{\gamma}^+ = \hat{\gamma}^+_1 \cdot
\hat{\gamma}^+_2$, a zero at $\infty^{\pm}$ and normalised by the
conditions $h_{\pm}(\infty^{\mp}) = 1$ respectively. From the
general theory of Riemann surfaces we can then write expressions
for these functions in terms of Riemann $\theta$-functions. This
relies on Riemann's theorem for the zeroes of a $\theta$-function,
which states (in the elliptic case $g = 1$) that if
$\theta(\mathcal{A}(P) - w)$ is not identically zero, then it has
a single zero $P_0$ (or a degree $g$ divisor of zeroes in the general
case) satisfying $\mathcal{A}(P_0) = w - \mathcal{K}$. So choosing $w
\equiv \mathcal{A}(Q_0) + \mathcal{K}$ yields $\mathcal{A}(P_0 \cdot
Q_0^{-1}) = 0$ which by Abel's theorem implies $P_0 = Q_0$. So we can
write $h_{\pm}$ as follows
\begin{equation*}
\begin{split}
h_-(P) &= \frac{\theta (\mathcal{A}(\infty^+) - w_1) \theta
(\mathcal{A}(\infty^+) - w_2)}{\theta (\mathcal{A}(\infty^+) -
w_-) \theta (\mathcal{A}(\infty^+) - w_0^-)} \frac{\theta
(\mathcal{A}(P) - w_-) \theta (\mathcal{A}(P) - w_0^-)}{\theta
(\mathcal{A}(P) -
w_1) \theta (\mathcal{A}(P) - w_2)}, \\
h_+(P) &= \frac{\theta (\mathcal{A}(\infty^-) - w_1) \theta
(\mathcal{A}(\infty^-) - w_2)}{\theta (\mathcal{A}(\infty^-) -
w_+) \theta (\mathcal{A}(\infty^-) - w_0^+)} \frac{\theta
(\mathcal{A}(P) - w_+) \theta (\mathcal{A}(P) - w_0^+)}{\theta
(\mathcal{A}(P) - w_1) \theta (\mathcal{A}(P) - w_2)},
\end{split}
\end{equation*}
where
\begin{equation*}
\begin{split}
w_i &\equiv \mathcal{A}(\hat{\gamma}^+_i) + \mathcal{K}, \qquad i = 1,2\\
w_{\pm} &\equiv \mathcal{A}(\infty^{\pm}) + \mathcal{K},\\
w_0^- &\equiv w_1 + w_2 - w_- = \mathcal{A}(\hat{\gamma}^+) -
\mathcal{A}(\infty^-) + \mathcal{K} = D,\\
w_0^+ &\equiv w_1 + w_2 - w_+ = \mathcal{A}(\hat{\gamma}^+) -
\mathcal{A}(\infty^+) + \mathcal{K} = D + \mathcal{A}(\infty^-).
\end{split}
\end{equation*}
One can absorb a lot of the common factors between $h_{\pm}(0^+)$
into the overall constant $C$ so that we are left with
\begin{equation} \label{h_pm(0)}
\begin{split}
h_-(0^+) &= \frac{\theta (\mathcal{A}(\infty^+) -
w_1) \theta (\mathcal{A}(\infty^+) - w_2)}{\theta
(\mathcal{A}(\infty^+) - w_-)} \frac{\theta (\mathcal{A}(0^+) -
w_0^-)}{\theta
(\mathcal{A}(0^+) - w_+)}\\
h_+(0^+) &= \frac{\theta (\mathcal{A}(\infty^-) - w_1) \theta
(\mathcal{A}(\infty^-) - w_2)}{\theta (\mathcal{A}(\infty^-) -
w_+)} \frac{\theta (\mathcal{A}(0^+) - w_0^+)}{\theta
(\mathcal{A}(0^+) - w_-)}.\\
\end{split}
\end{equation}
Likewise, the function $\chi(P)$ was defined as the unique
meromorphic function with zeroes at the divisor $\hat{\gamma}^+
\cdot \hat{\tau} \hat{\gamma}^+ = \hat{\gamma}^+_1 \cdot
\hat{\gamma}^+_2 \cdot \hat{\tau} \hat{\gamma}^+_1 \cdot
\hat{\tau} \hat{\gamma}^+_2$, poles at the branch points given by
the divisor $B = B_1 \cdot B_2 \cdot \hat{\tau} B_1 \cdot
\hat{\tau} B_2$ and normalised by the condition $\chi(\infty^+) =
1$, so that it can be constructed out of Riemann
$\theta$-functions as follows
\begin{multline*}
\chi(P) = \frac{\theta (\mathcal{A}(\infty^+) - b_1) \theta
(\mathcal{A}(\infty^+) - b_2) \theta (\mathcal{A}(\infty^+) +
\bar{b}_1) \theta (\mathcal{A}(\infty^+) + \bar{b}_2)}{\theta
(\mathcal{A}(\infty^+) - w_1) \theta (\mathcal{A}(\infty^+) - w_2)
\theta (\mathcal{A}(\infty^+) + \bar{w}_1) \theta
(\mathcal{A}(\infty^+) + \bar{w}_2)} \\ \times \frac{\theta
(\mathcal{A}(P) - w_1) \theta (\mathcal{A}(P) - w_2) \theta
(\mathcal{A}(P) + \bar{w}_1) \theta (\mathcal{A}(P) +
\bar{w}_2)}{\theta (\mathcal{A}(P) - b_1) \theta (\mathcal{A}(P) -
b_2) \theta (\mathcal{A}(P) + \bar{b}_1) \theta (\mathcal{A}(P) +
\bar{b}_2)}
\end{multline*}
where
\begin{equation*}
b_i \equiv \mathcal{A}(B_i) + \mathcal{K}, \quad i = 1,2
\end{equation*}
and we have used the fact that
\begin{equation*}
\begin{split}
- \bar{w}_i &= - \overline{\mathcal{A}(\hat{\gamma}^+_i)} -
\bar{\mathcal{K}},
= \mathcal{A}(\hat{\tau} \hat{\gamma}^+_i) + \mathcal{K}, \qquad
i = 1,2,\\
- \bar{b}_i &= - \overline{\mathcal{A}(B_i)} - \bar{\mathcal{K}},
= \mathcal{A}(\hat{\tau} B_i) + \mathcal{K}, \qquad i = 1,2.
\end{split}
\end{equation*}
We thus obtain the following expression for the real constant
$\chi(\infty^-)^{\frac{1}{2}}$,
\begin{equation} \label{chi(infty)}
\chi(\infty^-)^{\frac{1}{2}} = \frac{\left| \theta
(\mathcal{A}(\infty^+) - b_1) \theta (\mathcal{A}(\infty^+) - b_2)
\right|}{\left| \theta (\mathcal{A}(\infty^+) - w_1) \theta
(\mathcal{A}(\infty^+) - w_2) \right|} \frac{\left| \theta
(\mathcal{A}(\infty^-) - w_1) \theta (\mathcal{A}(\infty^-) - w_2)
\right|}{\left| \theta (\mathcal{A}(\infty^-) - b_1) \theta
(\mathcal{A}(\infty^-) - b_2) \right|}.
\end{equation}
Combining \eqref{h_pm(0)} and \eqref{chi(infty)} and absorbing
again some overall coefficient into the constant $C$ we obtain up
to a residual $SU(2)_R \times SU(2)_L$ transformation (i.e. up to
constant global phases)
\begin{equation*}
\begin{split}
h_-(0^+) &= \frac{\left| \theta (\mathcal{A}(\infty^+) - b_1)
\theta (\mathcal{A}(\infty^+) - b_2) \right|}{\theta
(\mathcal{A}(\infty^+) - w_-)} \frac{\theta (\mathcal{A}(0^+) -
w_0^-)}{\theta (\mathcal{A}(0^+) - w_+)}\\
\frac{h_+(0^+)}{\chi(\infty^-)^{\frac{1}{2}}} &= \frac{\left|
\theta (\mathcal{A}(\infty^-) - b_1) \theta (\mathcal{A}(\infty^-)
- b_2) \right|}{\theta (\mathcal{A}(\infty^-) - w_+)} \frac{\theta
(\mathcal{A}(0^+) - w_0^+)}{\theta (\mathcal{A}(0^+) - w_-)}.\\
\end{split}
\end{equation*}
This is as much as we can simply these normalisation constants without
explicitly evaluating them. To proceed we must now evaluate the arguments
of each $\theta$-function in turn.

\begin{figure}
\centering \psfrag{X1}{\footnotesize{$B_1$}}
\psfrag{X2}{\footnotesize{$B_2$}}
\psfrag{a}{\footnotesize{$a$}} \psfrag{b}{\footnotesize{$b$}}
\psfrag{c}{\footnotesize{$c$}}
\psfrag{c2}{\footnotesize{$\tilde{c}$}}
\psfrag{pinf}{\footnotesize{$\infty^+$}}
\psfrag{minf}{\footnotesize{$\infty^-$}}
\includegraphics[width=55mm]{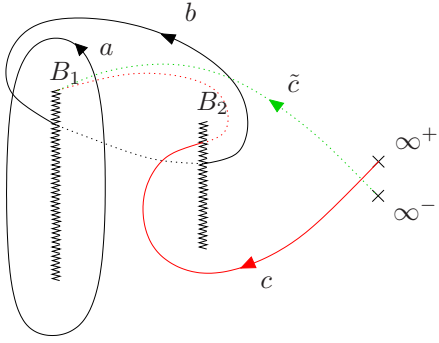}
\caption{Curves $c,\tilde{c}$ involved in computing $\mathcal{A}(B_1)$
and $\mathcal{A}(\infty^-) - \mathcal{A}(B_1)$. Notice that neither of
these two curves intersects the $a$- and $b$-cycles since they must lie
within $\Sigma_{\text{cut}}$.} \label{A(B_1)}
\end{figure}

\begin{figure}
\centering \psfrag{X1}{\footnotesize{$B_1$}}
\psfrag{X2}{\footnotesize{$B_2$}}
\psfrag{a}{\footnotesize{$a$}} \psfrag{b}{\footnotesize{$b$}}
\psfrag{c}{\footnotesize{$c_2$}}
\psfrag{c2}{\footnotesize{$\tilde{c}_2$}}
\psfrag{pinf}{\footnotesize{$\infty^+$}}
\psfrag{minf}{\footnotesize{$\infty^-$}}
\includegraphics[width=55mm]{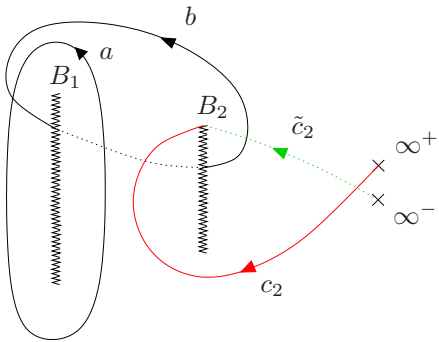}
\caption{Curves $c_2,\tilde{c}_2$ involved in computing $\mathcal{A}(B_2)$
and $\mathcal{A}(\infty^-) - \mathcal{A}(B_2)$. Again, neither of
these two curves intersects the $a$- and $b$-cycles since they must lie
within $\Sigma_{\text{cut}}$.} \label{A(B_2)}
\end{figure}

Now in the elliptic case there is just a single holomorphic $1$-form
$\omega$ so that the formula \eqref{K def} for $\mathcal{K}$ simply
yields $\mathcal{K} = 2 \pi \left( \frac{1}{2} + \frac{\tau}{2}
\right)$, and on the other hand we have $D = 2 \pi \left( \frac{1}{2}
+ X_0 \right)$. We will also need to make use of \eqref{rho def} for
the various integrals present. Using this information it is
straightforward to show that
\begin{equation} \label{theta relation 1}
\begin{split}
\theta \left( \mathcal{A}(0^+) - w_0^- \right) &= \vartheta_3 \left(
X_0 - i \rho_+ \right), \\
\theta \left( \mathcal{A}(0^+) - w_+ \right) &= \exp \left( -\pi i
\frac{\tau}{4} - \pi \rho_+ \right) \vartheta_2 \left( i \rho_+ \right), \\
\theta \left( \mathcal{A}(0^+) - w_0^+ \right) &= \exp \left( -\pi i
\frac{\tau}{4} + \pi \rho_- + \frac{i}{2} D \right)
\vartheta_1 \left( X_0 - i \rho_- \right), \\
\theta \left( \mathcal{A}(0^+) - w_- \right) &= \vartheta_0 \left( i
\rho_- \right).
\end{split}
\end{equation}
Next, one has $\mathcal{A}(\infty^+) - w_- = - (\mathcal{A}(\infty^-)
+ \mathcal{K})$ and $\mathcal{A}(\infty^-) - w_+ =
\mathcal{A}(\infty^-) - \mathcal{K}$ where $\mathcal{A}(\infty^-)$ can
be evaluated in terms of \eqref{rho def} using \eqref{int rel}
yielding $\mathcal{A}(\infty^-) = 2 \pi \left( \frac{1 - \tau}{2} +
i(\rho_+ - \rho_-) \right)$. This immediately leads to
\begin{equation} \label{theta relation 2}
\theta \left( \mathcal{A}(\infty^+) - w_- \right) = \exp \left( \pi i
\tau + 2 \pi (\rho_+ - \rho_-) \right) \theta \left(
\mathcal{A}(\infty^-) - w_+ \right).
\end{equation}
Finally we must obtain relations among the $\theta$-functions
involving branch points. We have for instance
\begin{equation*}
\mathcal{A}(B_1) = 2 \pi \int_c \omega, \qquad \mathcal{A}(\infty^-) -
\mathcal{A}(B_1) = - 2 \pi \int_{\tilde{c}} \omega,
\end{equation*}
where the curves $c,\tilde{c}$ are depicted in Figure \ref{A(B_1)}.
Recall that we had chosen a branch for the Abel integral
$\mathcal{A}(P) = 2 \pi \int_{\infty^+}^P \omega$ by requiring that
the path from $\infty^+$ to $P$ stayed within
$\Sigma_{\text{cut}}$. This is insured by taking paths $c$ that respect
the following intersection numbers
\begin{equation*}
c \cap a = c \cap b = 0.
\end{equation*}
The curves $c, \tilde{c}$ are related by $\hat{\sigma} c - b + a =
\tilde{c}$, which leads to (using also $\hat{\sigma}^{\ast} \omega =
- \omega$)
\begin{equation} \label{theta relation 3}
\theta \left( \mathcal{A}(\infty^+) - b_1 \right) = \theta \left(
\mathcal{A}(\infty^-) - b_1 \right).
\end{equation}
For the other branch point $B_2$ things are very similar. In
particular we have again
\begin{equation*}
\mathcal{A}(B_2) = 2 \pi \int_{c_2} \omega, \qquad \mathcal{A}(\infty^-) -
\mathcal{A}(B_2) = - 2 \pi \int_{\tilde{c}_2} \omega,
\end{equation*}
where the curves $c_2,\tilde{c}_2$ are depicted in Figure \ref{A(B_2)}.
However, the curves $c_2, \tilde{c}_2$ are now related by
$\hat{\sigma} c_2 + a = \tilde{c}_2$, which leads to
\begin{equation} \label{theta relation 4}
| \theta \left( \mathcal{A}(\infty^+) - b_2 \right)| = \exp \left( \pi
i \frac{\tau}{2} + \pi (\rho_+ - \rho_-) \right) | \theta \left(
\mathcal{A}(\infty^-) - b_2 \right)|.
\end{equation}

So at last, gathering together equations \eqref{theta relation 1}, \eqref{theta
relation 2}, \eqref{theta relation 3}, \eqref{theta relation 4} and
applying a global $SU(2)_R \times SU(2)_L$ transformation to remove a
relative phase $\exp \left( \frac{i}{2} D \right)$ appearing in
$h_+(0^+)$ we obtain the expression \eqref{prefactors} for the
normalisation constants.

\end{document}